\documentclass[prl,aps,twocolumn,superscriptaddress,floatfix,showpacs,longbibliography]{revtex4-2}
\usepackage{graphicx}
\usepackage{dcolumn}
\usepackage{bm}
\usepackage{upgreek}
\usepackage{natbib}
\usepackage[normalem]{ulem}
\usepackage{amsmath,amssymb}
\usepackage{dsfont}
\usepackage[english]{babel}
\usepackage{color}

\begin{document}

\title{Searching for Unconventional Superfluid in Excitons of Monolayer
Semiconductors}

\author{Wei Chen}
\affiliation{Guangdong Basic Research Center of Excellence for Structure and Fundamental Interactions of Matter, Guangdong Provincial Key Laboratory of Quantum Engineering and Quantum Materials, School of Physics, South China Normal University, Guangzhou 510006, China}
\affiliation{Guangdong-Hong Kong Joint Laboratory of Quantum Matter, Frontier
Research Institute for Physics, South China Normal University, Guangzhou
510006, China}
\author{Chun-Jiong Huang}
\affiliation{Department of Physics and HKU-UCAS Joint Institute for
Theoretical and Computational Physics at Hong Kong, The University of Hong
Kong, Hong Kong, China}
\author{Qizhong Zhu}
\email{qzzhu@m.scnu.edu.cn}
\affiliation{Guangdong Basic Research Center of Excellence for Structure and Fundamental Interactions of Matter, Guangdong Provincial Key Laboratory of Quantum Engineering and Quantum Materials, School of Physics, South China Normal University, Guangzhou 510006, China}
\affiliation{Guangdong-Hong Kong Joint Laboratory of Quantum Matter, Frontier
Research Institute for Physics, South China Normal University, Guangzhou
510006, China}

\date{\today}

\begin{abstract}
  It is well known that two-dimensional (2D) bosons in homogeneous space cannot undergo real
  Bose-Einstein condensation, and the superfluid to normal phase transition is
  Berezinskii-Kosterlitz-Thouless (BKT) type, associated with vortex-antivortex pair unbinding. 
  Here we point out a 2D bosonic system whose low energy physics goes
   beyond conventional paradigm of 2D {\it homogeneous} bosons, i.e.,
  intralayer excitons in monolayer transition metal dichalcogenides.
  With intrinsic valley-orbit coupling and valley Zeeman energy, 
exciton dispersion becomes linear at small momentum, giving rise to a series of novel features.  
  The critical temperature of Bose-Einstein condensation of these excitons is nonzero, suggesting
  true long-range order in 2D homogeneous system.
  The dispersion of Goldstone mode at
  long wavelength has the form $\varepsilon(\boldsymbol{q})\sim\sqrt{q}$, in contrast to
  conventional linear phonon spectrum. The vortex energy
  deviates from the usual logarithmic form with respect to system size, but instead has an additional linear term. 
  Superfluid to normal phase transition is no longer BKT type for system size beyond a
  characteristic scale, without discontinuous jump in superfluid density.
  With the recent experimental progress on exciton fluid at thermal equilibrium in monolayer semiconductors,
  our work points out an experimentally accessible system to search for unconventional 2D superfluids beyond BKT paradigm.
\end{abstract}

{\maketitle}

In two-dimensional (2D) homogeneous systems, it is well known that continuous symmetry cannot
be broken spontaneously according to Mermin-Wagner theorem \cite{mermin_Absence_1966,hohenberg_Existence_1967}, and there is no true long-range order. As a special
example, Bose-Einstein condensation (BEC) critical temperature in 2D is zero. Nevertheless, superfluid is
still possible at finite temperature, and the transition from superfluid to
normal phase is described by the Berezinskii-Kosterlitz-Thouless (BKT) theory \cite{berezinsky_DESTRUCTION_1972,kosterlitz_Ordering_1973}, where the underlying mechanism is the
vortex-antivortex pair unbinding at high temperature. This generic paradigm is successful in the description of a variety of 2D superfluids,
including liquid helium films \cite{bishop_Study_1978}, superconductors \cite{resnick_KosterlitzThouless_1981}, cold atomic gases \cite{hadzibabic_Berezinskii_2006,murthy_Observation_2015}, exciton-polariton condensates \cite{caputo_Topological_2018}, and
dipolar excitons \cite{dang_Observation_2020}.

Generically, for a 2D bosonic system, the nature of low temperature phases
depends crucially on the density of states. Deviation from homogeneous space or parabolic dispersion may lead to superfluids beyond the conventional paradigm.
For example, when ultracold bosonic atoms are confined in a harmonic trap, the BEC critical temperature is
nonzero \cite{dalfovo_Theory_1999}. Similar examples include bosons confined
on the surface of a sphere, where a finite BEC critical temperature also exists due to finite size effect \cite{tononi_BoseEinstein_2019}. 
On the other hand, for homogeneous bosons with quartic
dispersion realized at the transition point of spin-orbit coupled 
gases, BKT transition temperature vanishes and the low temperature phase is
characterized by an algebraic order \cite{po_Twodimensional_2015}. 
This is an interesting example of enhanced low energy fluctuations brought by increased density of states.
The contrary case, i.e., interacting bosons with single-particle dispersion $\epsilon(k)\sim k^{\nu}$ ($\nu<2$) in realistic experimental systems are still lacking. This type of system is of fundamental importance and interesting, as one of the three exhaustive cases of an isotropic 2D bosonic system in {\it homogeneous} space, e.g.,
$\nu<2$, $\nu=2$, and $\nu>2$.
In a word,
2D {\it homogeneous} bosons beyond the conventional paradigm are quite rare and highly interesting, and will surely enrich our understanding of such fundamental concepts as BEC and superfluidity. 

In recent years, there has been growing interest in the realization of exciton condensation
in bilayer 2D materials \cite{berman_Superfluidity_2012,
fogler_Hightemperature_2014,wu_Theory_2015,berman_Hightemperature_2016,berman_Superfluidity_2017,
debnath_Exciton_2017,
zhu_Gate_2019,wang_Evidence_2019,ma_Strongly_2021,remez_Leaky_2022,deng_Moir_2022,zimmerman_Collective_2022}, such as transition metal dichalcogenide (TMD).  
Besides the long-sought interlayer exciton condensation in heterobilayers, the possibility of intralayer exciton condensation in monolayer TMD
can not be excluded either \cite{guo_Tuning_2022}, despite of its relatively short lifetime.
In particular, significant experimental progress has been achieved 
in the exciton fluid at thermal equilibrium in monolayer TMD \cite{delaguila_Ultrafast_2023}, which paves the way towards the realization of exciton superfluid in this system.

\begin{figure}
\centering
\includegraphics[width=0.99\linewidth]{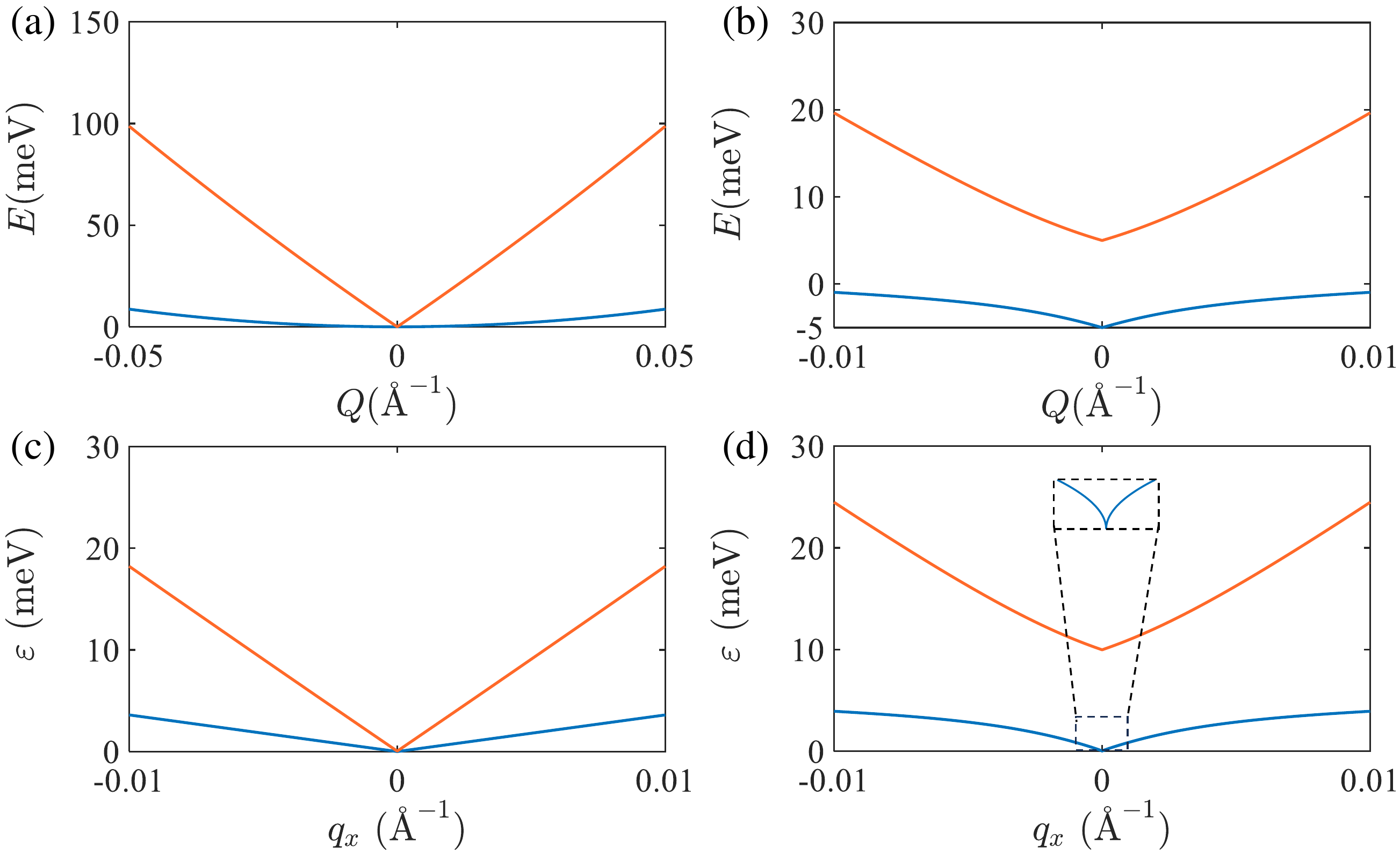}
\caption{Single exciton dispersion for $\delta=0$ (a) and $\delta=5$ meV (b).
Bogoliubov excitation spectra of an exciton condensate at zero momentum for $\delta=0$ (c) and $\delta=5$ meV (d). 
Inset of (d) shows the zoom-in plot of dispersion $\varepsilon(\boldsymbol{q})\sim\sqrt{q}$. All spectra in (a)-(d)
have rotational symmetry in the 2D $\boldsymbol{Q}$ or $\boldsymbol{q}$ plane. Unless otherwise specified,
the exciton-exciton interaction strengths
chosen throughout this paper are
$c_1=1.0$ eV$\cdot$nm$^2$, $c_2=0.6$ eV$\cdot$nm$^2$, within the same order of magnitude as the values calculated in Ref. \cite{shahnazaryan_Excitonexciton_2017,erkensten_Excitonexciton_2021}. $n_0=9.9*10^9$ cm$^{-2}$.
 \label{fig1}}
\end{figure}

Here we point out that in the system of monolayer TMD,
intralayer exciton with linear dispersion \cite{yu_Dirac_2014,qiu_Nonanalyticity_2015} provides another example beyond conventional paradigm of superfluids.
Note that system anisotropy can lead to interesting anisotropic superfluids in 2D materials \cite{berman_BoseEinstein_2017,saberi-pouya_Hightemperature_2018,kezerashvili_Superfluidity_2022}, also beyond the conventional paradigm of isotropic superfluids, but here we only consider the isotropic superfluids.
The center-of-mass motion of intralayer exciton in monolayer TMD features intrinsic valley-orbit coupling \cite{yu_Dirac_2014,qiu_Nonanalyticity_2015},
originating from electron-hole exchange interaction. With additional valley Zeeman energy introduced by either magnetic field or valley-selective optical Stark effect, the dispersion of intralayer exciton at lower branch 
becomes linear (see Fig. \ref{fig1}(b)). We will reveal that this special dispersion endows intralayer excitons with
a variety of novel features in low energy physics. Our main findings
include the following: (i) With a valley Zeeman energy, the BEC critical
temperature of this 2D {\it homogeneous} system is nonzero and exhibits rapid increase with the
field strength. (ii) The Bogoliubov
excitation spectrum of this exciton condensate shows $\varepsilon(\boldsymbol{q})\sim\sqrt{q}$
behaviour at long wavelength in the presence of valley Zeeman energy, which is an usual form of gapless Goldstone mode,
in contrast to conventional phonon excitation. (iii) The vortex energy deviates from logarithmic form 
with respect to system size, resulting in a non-BKT-type
phase transition for large system size. There exists a characteristic system
size, beyond which the superfluid to normal phase transition evolves from BKT type to 
3D-like without discontinuous jump in superfluid density. 
This crossover can also be observed in a single system by tuning the magnitude of valley Zeeman energy. 
(iv) With the increase of temperature, the system undergoes a two-step phase transition,
first from a BEC with long-range order to a superfluid with quasi-long-range
 order, and then to a normal phase. 
These novel phases can be experimentally
detected by measuring the spatial and temporal coherence of emitted photons by excitons.

\begin{figure}
\centering
\includegraphics[width=0.99\linewidth]{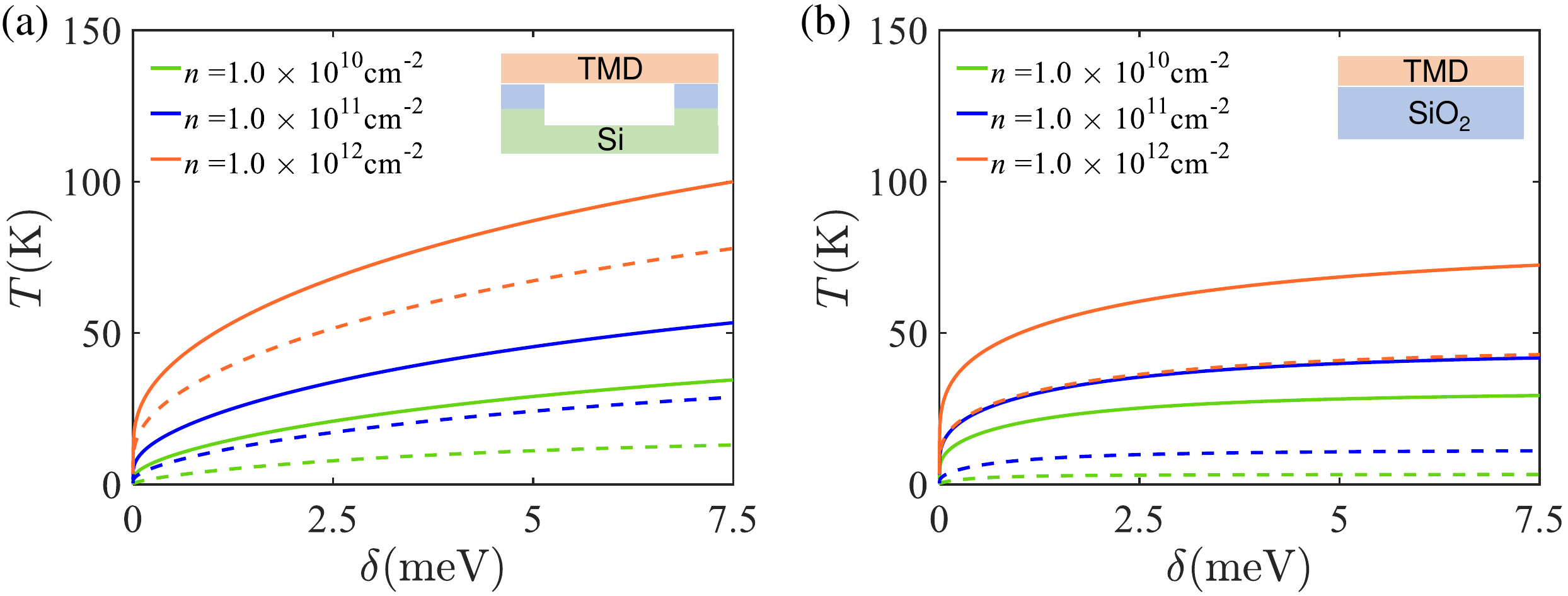}
\caption{Change of non-interacting (interacting) BEC critical temperature with $\delta$ shown in dashed (solid) line at different exciton densities, in the free-standing case \cite{sun_Enhanced_2022} (a) and for a TMD placed on SiO$_2$ substrate (b), with schematics shown in the insets. 
Lines with same color correspond to same exciton densities. The relations $A=0.9$ eV$\cdot\mathrm{\AA}/\epsilon_d^2$ and $r_0=33.875$\AA$/\epsilon_d$ are used here \cite{Supplement_Material}, with $\epsilon_d=1$ in (a) and $\epsilon_d=2.5$ in (b). The screened exciton-exciton interaction constants are estimated by assuming the $1/\epsilon_d$ dependence. The maximum $\delta=7.5$meV corresponds to a magnetic field of $\sim$65T, achievable in existing experiments \cite{stier_Magnetooptics_2018,liu_Magnetophotoluminescence_2019,goryca_Revealing_2019}. \label{fig2}}
\end{figure}

\begin{figure}
\centering
\includegraphics[width=0.99\linewidth]{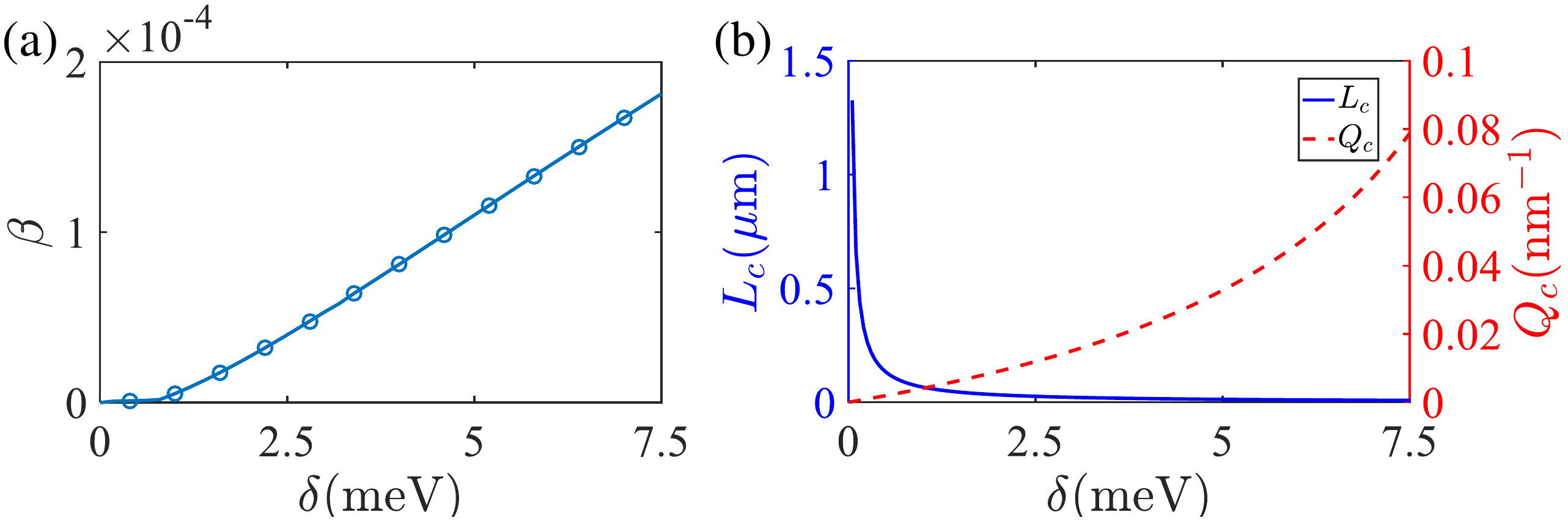}
\caption{(a) Change of coefficient $\beta$ with $\delta$. (b) Characteristic system size $L_c$ and momentum $Q_c$ as functions of $\delta$. \label{fig3}}
\end{figure}

{\textit{Monolayer exciton dispersion}}. There are two inequivalent valleys in the Brillouin zone corner of monolayer TMD, 
denoted as $\pm K$ valley, respectively.
Valley excitons in monolayer TMD behave as a
pseudospin-$1/2$ bosonic system, whose valley pseudospin is coupled with
center-of-mass momentum of excitons \cite{yu_Dirac_2014,qiu_Nonanalyticity_2015}, giving rise to the valley-orbit coupling.
The effective Hamiltonian describing the center-of-mass motion of intralayer excitons with valley-orbit coupling reads \cite{yu_Dirac_2014,wu_Exciton_2015,qiu_Nonanalyticity_2015,wu_Topological_2017,sauer_Optical_2021,salvador_Optical_2022}
\begin{align}
 \hat{H}_0 = & \frac{\hbar^2 Q^2}{2 m} + A Q+ A Q \cos (2
\theta_{\boldsymbol{Q}})\sigma_x + A Q \sin (2 \theta_{\boldsymbol{Q}})\sigma_y \nonumber\\
& + \delta \sigma_z.
\end{align}
Here $\boldsymbol{Q}$ is the center-of-mass momentum of exciton with magnitude $Q=|\boldsymbol{Q}|$ and angle $\theta_{\boldsymbol{Q}}$, $m\approx1.1m_0$ is exciton effective mass, $\sigma_i$ $(i=x, y, z)$ are
the Pauli matrices of valley pseudospin, and $A\approx0.9$ eV$\cdot\mathrm{\AA}$  is the valley-orbit
coupling strength \cite{qiu_Nonanalyticity_2015}, related with electron-hole exchange interaction.
With the exchange interaction modeled by the screened Keldysh potential \cite{wu_Exciton_2015,sauer_Optical_2021}, 
which gives a good description of the exciton Rydberg series in monolayer TMD \cite{chernikov_Exciton_2014,stier_Magnetooptics_2018},
the $AQ$ term is an approximation of the more accurate $AQ/(1+r_0 Q)$ term, where both $A$ and the screening length $r_0$ depend on the effective dielectric constant $\epsilon_d$, determined by surrounding environment. Unless otherwise specified, the value of $A$ adopted here corresponds to a free-standing TMD monolayer with $r_0$ neglected, realizable with TMD placed on top of circular holes prepatterned on
SiO$_2$ substrate, using the experimental setup in Ref. \cite{sun_Enhanced_2022}. For TMD placed on substrate materials, the screening
will reduce the value of $A$ and hence modifies the quantitative results such as the critical temperature calculated below (see Fig. \ref{fig2} and Ref. \cite{Supplement_Material}). We also introduce a
$\delta\sigma_z$ term, known as the valley Zeeman energy, with magnitude $\delta$ tunable by applying a magnetic field \cite{aivazian_Magnetic_2015,stier_Exciton_2016} or utilizing the valley-selective optical Stark effect \cite{kim_Ultrafast_2014,sie_Valleyselective_2015}.
The non-interacting
dispersion has two branches, given by $\xi_{\pm}(\boldsymbol{Q})=\hbar^2 Q^2/2m+A Q\pm\sqrt{A^2 Q^2+\delta^2}$, as shown in Figs. \ref{fig1}(a) and (b). 
For finite $\delta$, the lower branch of dispersion at small momentum features a linear spectrum, distinct from usual parabolic dispersion
and brings
dramatic change to low energy properties of this system. The range of
linear dispersion defines a characteristic momentum $Q_c$, at which the parabolic and linear terms are comparable in weight, and corresponds to a characteristic system size $L_c=1/Q_c$, both shown in Fig. \ref{fig3}(b).

{\textit{Non-interacting BEC critical temperature}}. In the absence of valley Zeeman energy, the lower branch of dispersion is parabolic,
and obviously the BEC critical temperature vanishes, as in conventional 2D bosonic systems.
With valley Zeeman energy ($\delta\neq0$), the lower branch of dispersion is linear
at small momentum, which modifies the low energy density of
states and renders a BEC at finite temperature possible. We first neglect
the exciton-exciton interaction and calculate the non-interacting condensation temperature $T_\mathrm{BEC}$
through the relation,
\begin{equation}
n=\frac{1}{(2\pi)^2}\int\sum_{\tau=\pm}\frac{d^2\boldsymbol{Q}}{e^{\beta[\xi_\tau
   (\boldsymbol{Q})-\mu(T)]}-1},
\end{equation}
where $\mu$ is the exciton chemical potential and $\beta=1/k_B T$.
When $T\rightarrow T_\mathrm{BEC}$, $\mu\rightarrow -\delta$. The calculated relation between $T_\mathrm{BEC}$ and $\delta$ is shown in Fig. \ref{fig2} for
different exciton densities in the two cases illustrated by the insets. Clearly, $T_\mathrm{BEC}$ monotonically increases with
$\delta$. For typical exciton density $n= 10^{10}\sim
10^{12}$ cm$^{-2}$ below the Mott limit ($\sim
10^{13}$ cm$^{-2}$) \cite{delaguila_Ultrafast_2023}, a moderate $\delta$ can lead to a relatively high
$T_\mathrm{BEC}$ on the order of $10\sim100$ K in the free-standing case. This is in stark contrast to the conventional bosonic system with parabolic dispersion, 
where $T_\mathrm{BEC}=0$ and only 
a quasi-condensate exists at finite temperature. Here one has
a true 2D condensate with long-range order, in homogeneous space. 
As will be shown below, we find that the inclusion of exciton-exciton interaction can further enhance the BEC critical temperature.

{\textit{Bogoliubov excitation spectrum.}} The linear dispersion of
non-interacting exciton at small momentum also implies unusual 
low energy excitations of exciton condensate. Assuming that excitons condense at 
zero momentum state, the low energy excitation can be calculated by the standard
Bogoliubov theory (see for example, \cite{pethick_Bose_2008}). The mean-field energy functional reads
\begin{align}
		E[\Psi_\sigma] &= \int d^2\boldsymbol{r} \Bigg\{\left(\Psi_\uparrow^*,\Psi_\downarrow^*\right)\Big[\left(\frac{\hbar^2 {Q}^2}{2m}+A Q\right)+\delta\sigma_z \nonumber \\
		&+A Q \cos(2\theta_{\boldsymbol{Q}})\sigma_x+A Q\sin(2\theta_{\boldsymbol{Q}})\sigma_y \Big]
		\left( \begin{array}{r}
			\Psi_\uparrow  \\
			\Psi_\downarrow
		\end{array}  \right)\nonumber\\
	& + \frac{c_1}{2}\left(\left|\Psi_\uparrow\right|^4+\left|\Psi_\downarrow\right|^4\right)+{{c}_2}{\left|{{\Psi_{\uparrow}}} \right|^2}{\left| {{\Psi_{\downarrow}}} \right|^2}\Bigg\},
	\label{mfen}
\end{align}
where $\Psi_\sigma$ is the condensate wave function for pseudospin $\sigma=\uparrow,\downarrow$, corresponding to $\pm K$ valley. At the qualitative level, the exciton-exciton interaction is modelled as a contact interaction, which is valid when the exciton density is low \cite{ben-taboude-leon_Excitonexciton_2001,berman_Hightemperature_2016}, with $c_1$ and $c_2$ being exciton-exciton interaction strengths
between the same and different pseudospin states, respectively.

Within the same framework, the condensate dynamics can be described by the
spinor Gross-Pitaevskii (GP) equation,
\begin{equation}	
i\hbar \frac{\partial }{{\partial t}}\left( {\begin{array}{*{20}{c}}
		{{\Psi_\uparrow}}  \\
		{{\Psi_\downarrow}}  \\
\end{array}} \right) = \left( {\begin{array}{*{20}{c}}
		{H_{\uparrow\uparrow}} & {{H_A}}  \\
		{H_A^*} & {H_{\downarrow\downarrow}}  \\
\end{array}} \right)\left( {\begin{array}{*{20}{c}}
		{{\Psi_\uparrow}}  \\
		{{\Psi_\downarrow}}  \\
\end{array}} \right),
\end{equation}
where $H_{\uparrow\uparrow} = \hbar^2 Q^2 / 2 m + A {Q} + \delta+ {c_1}{{\left| {{\Psi_\uparrow}} \right|}^2} + {c_2}{{\left| {{\Psi_\downarrow}} \right|}^2}$, $H_{\downarrow\downarrow} = \hbar^2 Q^2 / 2 m + A {Q} -\delta+ {c_1}{{\left| {{\Psi_\downarrow}} \right|}^2}+ {c_2}{{\left| {{\Psi_\uparrow}} \right|}^2} $, and
$H_A = AQ\exp(-2i\theta_{\boldsymbol{Q}})$. 
In the presence of valley Zeeman energy ($\delta>0$), minimization of the mean-field energy functional (Eq. \ref{mfen}) gives the wave function of an exciton condensate at ground state. 
The ground state is found to be $(|\Psi_{\uparrow}|^2,|\Psi_{\downarrow}|^2)=n_0(1/2-\delta/(n_0c_1-n_0c_2),
1/2+\delta/(n_0c_1-n_0c_2))$ for $0\leq\delta\leq(n_0c_1-n_0c_2)/2$, while $(|\Psi_{\uparrow}|^2,|\Psi_{\downarrow}|^2)=n_0(0,1)$ for
$\delta>(n_0c_1-n_0c_2)/2$,
with $n_0$ being the condensate density. 
For intralayer exciton, $c_1>c_2$ ensures that when $\delta=0$, exciton densities in two valleys are equal at ground state.
With the estimated values $c_1\sim1.0$ eV$\cdot$nm$^2$, $c_2\sim0.6$ eV$\cdot$nm$^2$ \cite{shahnazaryan_Excitonexciton_2017,erkensten_Excitonexciton_2021}, 
and typical exciton density $n_0=1.0*10^{11}$ cm$^{-2}$, the critical $\delta_c\equiv(n_0c_1-n_0c_2)/2\sim0.2$ meV,
beyond which the excitons are completely valley polarized. In the following, we consider the case $\delta>\delta_c$,
which is readily accessible in experiment with moderate magnetic field and greatly simplifies the calculations. Results within the
small range $0<\delta<\delta_c$ are also calculated with the wave function $n_0(0,1)$ as an approximation.

By expanding the wave function around the stationary state, $\Psi'_\sigma = \Psi_\sigma + \delta
\Psi_\sigma$, and assuming the form of perturbation $\delta\Psi_\sigma = \Psi_\sigma e^{- i \mu /
\hbar} [u_\sigma(\boldsymbol{q}) e^{- i \varepsilon(\boldsymbol{q}) t} - v_\sigma^{\ast} (\boldsymbol{q}) e^{i \varepsilon(\boldsymbol{q}) t}]$, 
one arrives at the
Bogoliubov equation for the low energy excitation,
\begin{eqnarray}
  \mathcal{M} \left( \begin{array}{c}
    u_\uparrow\\
    u_\downarrow\\
    v_\uparrow\\
    v_\downarrow
  \end{array} \right) = \varepsilon \left( \begin{array}{c}
    u_\uparrow\\
    u_\downarrow\\
    v_\uparrow\\
    v_\downarrow
  \end{array} \right),
\end{eqnarray}
with
\begin{eqnarray}
	\mathcal{M}=
	\begin{pmatrix}
		H_1^+ & B^+_1 & 0 & 0\\
		B^+_2 & H_2^+ & 0 &-n_0 c_1\\
		0 & 0 & H_1^- & B^-_1\\
		0 & n_0 c_1 & B^-_2 & H_2^-\\
	\end{pmatrix}.
\end{eqnarray}
Here $H_1^{\pm} =\pm\left(\hbar^2 q^2/{2m} + A q +
{n_0 c_2} - \mu+\delta\right)$, $H_2^{\pm} =\pm\left(\hbar^2 q^2/{2m} + A q +
{2}{n_0 c_1} - \mu -\delta\right)$,
$B^{\pm}_1= \pm A\left(q^2_x - q_y^2- 2i q_x q_y\right)/q
 $, $B^{\pm}_2= \pm A\left(q^2_x - q_y^2+ 2i q_x q_y\right)/q
$, and  $\mu= -\delta+n_0 c_1 $.
There are four groups of eigenvalues and only the two whose
corresponding eigenvectors satisfy $| u_\sigma |^2 - | v_\sigma |^2 = 1$ are
physical.
The two branches of excitations are
\begin{equation}
\varepsilon_{\pm}(\boldsymbol{q})=\frac{1}{2}\sqrt{\mathcal{A} + \mathcal{B} \pm 2\sqrt{\mathcal{C}^2- \mathcal{D}}},
\end{equation}
where the expressions of $\mathcal{A}$, $\mathcal{B}$, $\mathcal{C}$ and $\mathcal{D}$ are lengthy and listed in the
Supplementary Material \cite{Supplement_Material}. As shown in Figs. \ref{fig1}(c) and \ref{fig1}(d), for $\delta=0$, $\varepsilon_-(\boldsymbol{q})\sim q$ at small $q$,
while $\varepsilon_-(\boldsymbol{q})\sim\sqrt{q}$ for $\delta>0$ \cite{Supplement_Material}, which is a new form of gapless Goldstone mode unreported before.

{\textit{Interacting BEC critical temperature.}} With exciton-exciton interaction, the dispersion of low energy excitation is modified,
and the resulting change of density of states also affects the BEC critical temperature.
We quantitatively calculate the interacting critical temperature using the 
standard Hartree-Fock-Bogoliubov-Popov theory \cite{griffin_Conserving_1996,hutchinson_Finite_1997,shi_Finitetemperature_1998},
by solving self-consistently the
total exciton density \cite{Supplement_Material}
\begin{equation}
n =   n_0+\sum_{{\tau=\pm}\atop{\sigma=\uparrow,\downarrow}}\int\frac{d^2\boldsymbol{q}}{(2\pi)^2}\left\{|v_{\sigma}^{\tau}(\boldsymbol{q})|^2 
 +\frac{|u_{\sigma}^{\tau}(\boldsymbol{q})|^2+|v_{\sigma}^{\tau}(\boldsymbol{q})|^2}
{e^{\varepsilon_{\tau}(\boldsymbol{q})/k_B T}-1}\right\}, 
\end{equation}
where the condensate density $n_0$ also enters the excitation spectrum and quasiparticle amplitudes
 $u_{\sigma}(\boldsymbol{q})$ and $v_{\sigma}(\boldsymbol{q})$. By calculating the dependence of $n_0$ on $T$, one can extrapolate to
$n_0\rightarrow0$ and find the critical temperature $T_\mathrm{BEC}$ with interaction.
As shown in Fig. \ref{fig2}, the BEC critical temperature is enhanced by exciton-exciton interaction, which can be qualitatively understood
by considering that the density of states is reduced at low energy and thereby the condensate fraction at given
temperature is increased compared with non-interacting case.

{\textit{Vortex energy.}} In 2D bosonic systems, besides the non-singular excitations calculated above, there are also singular
topological excitations, i.e., vortices, which play a decisive role in conventional BKT theory.
In a single-component condensate, the appearance of a free vortex causes
an energy increase $E_\mathrm{v}\sim m n_s/2\int d\boldsymbol{r} v^2(\boldsymbol{r})\simeq\pi\hbar^2 n_s/m\ln (L/\zeta)$, where $v(\boldsymbol{r})$ is
the magnitude of velocity, $L$ is the system size,
$\zeta=\sqrt{\hbar^2/2m n_s c_1}$ is the condensate healing length, and $n_s$ is the superfluid density. This energy function has the same form with increase of entropy
$S\simeq 2 k_B \ln (L/\zeta)$ associated with a free vortex, thereby leading to the free energy change $\Delta\mathcal{F}=E_\mathrm{v}-TS\simeq( \pi\hbar^2 n_s/m-2k_B T)\ln (L/\zeta)$,
whose turning point gives $T_\mathrm{SF}$.
This is the case for parabolic particle dispersion, and since intralayer exciton exhibits different dispersion here,
the vortex energy is modified, as well as the nature of superfluid phase transition. 
In this pseudospin-1/2 system, vortices generally have two components with respective vorticity or circulation.
Here we are interested in the case of $\delta>\delta_c$, where the condensate at ground state is pseudospin polarized, and thus we make the approximation by
considering a single-component vortex with modified dispersion, i.e.,
$\xi_-(\boldsymbol{Q}) = \hbar^2 Q^2 / 2 m + A Q -\sqrt{A^2 Q^2 + \delta^2}$.
With similar argument, the vortex energy should be modified as $E_\mathrm{v}\sim m n_s/2\int d\boldsymbol{r} \left\{v^2(\boldsymbol{r})+(mA/\hbar) v(\boldsymbol{r})\right\}
\sim (\pi\hbar^2 n_s/m) \left[\alpha\ln (L/\zeta)+\beta (L/\zeta)\right]$,
where $\alpha$ and $\beta$ are two constants dependent on $A$ and $\delta$. In particular, $\beta$ should vanish at $\delta=0$ and increase with $\delta$.
Similar form of vortex energy is found
in superconductors or two-component BECs with Josephson coupling \cite{cataudella_Simple_1990,jensen_Twodimensional_1991,miu_Lengthscaledependent_1998,babaev_Phase_2004,kobayashi_BerezinskiiKosterlitzThouless_2019},
and the additional linear term $\beta (L/\zeta)$ implies the breakdown of BKT theory for large system size.

To be on a firmer ground, we adopt the trial wave function of a vortex as in the single-component case, i.e.,
 $\psi(\boldsymbol{r})\approx \sqrt{n_s} r e^{i\phi_{\boldsymbol{r}}}/{\sqrt{r^2 + 2}}$, where $r$ is in unit of $\zeta$, being a good fit to the numerical solution by Gross-Pitaevskii equation \cite{fetter_Rotating_2009}.
We numerically calculate the kinetic energy increase associated with the vortex within the new
dispersion, through
$E_\mathrm{v}\simeq\int d\boldsymbol{r}\psi^*(\boldsymbol{r})\hat{H}_\mathrm{kin}\psi(\boldsymbol{r})$, with $\hat{H}_\mathrm{kin}$ being the
kinetic energy operator corresponding to dispersion $\xi_-(\boldsymbol{Q})$.
The operation is carried out in momentum space,
bypassing the difficulty in expressing the valley-orbit coupling term in coordinate space \cite{Supplement_Material}.
 The obtained dependence of $E_\mathrm{v}$ on $L/\zeta$ can be fitted well with the relation above,
 which gives the specific value of $\alpha$ and $\beta$. Within the experimentally feasible range of
 $\delta\lesssim7.5$ meV, we find $\alpha\approx1$ and the change of $\beta$ with $\delta$ is shown in Fig. \ref{fig3}(a). The condition of equal contribution
 from these two terms $\alpha\ln (L/\zeta)+\beta (L/\zeta)$ also determines a characteristic system size, in quantitative agreement with $L_c$
 defined above.

{\textit{Superfluid to normal phase transition}}.
The linear term of vortex energy will dominate for large system size, and since this term grows faster than entropy, 
proliferation of free vortices will always be suppressed even at high temperature.
 This means the main contribution to the
depletion of superfluid density will come from non-singular excitations, including the Bogoliubov excitation with $\sqrt{q}$ dispersion,
and vortex-antivortex bound pairs.
These types of excitations will lead to continuous decrease of superfluid density with temperature, without
discontinuous jump in superfluid density at the phase transition point, similar to 3D case.
So we conclude that, for system size beyond a characteristic scale $L_c$, the superfluid
phase transition is no longer BKT type. On the other hand, for small system size with large lower bound of momentum, 
the effect of linear dispersion will be minor, and in this case we still expect a BKT-type finite-size crossover.
Note also that the characteristic system size $L_c$ depends on $\delta$, and thus the two limiting case above can also be
observed in a single system by tuning $\delta$. For small $\delta$, $L_c$ is on the order of $\mu$m, fully within the range of sample size
in current experiments.

\begin{figure}
\centering
\includegraphics[width=0.7\linewidth]{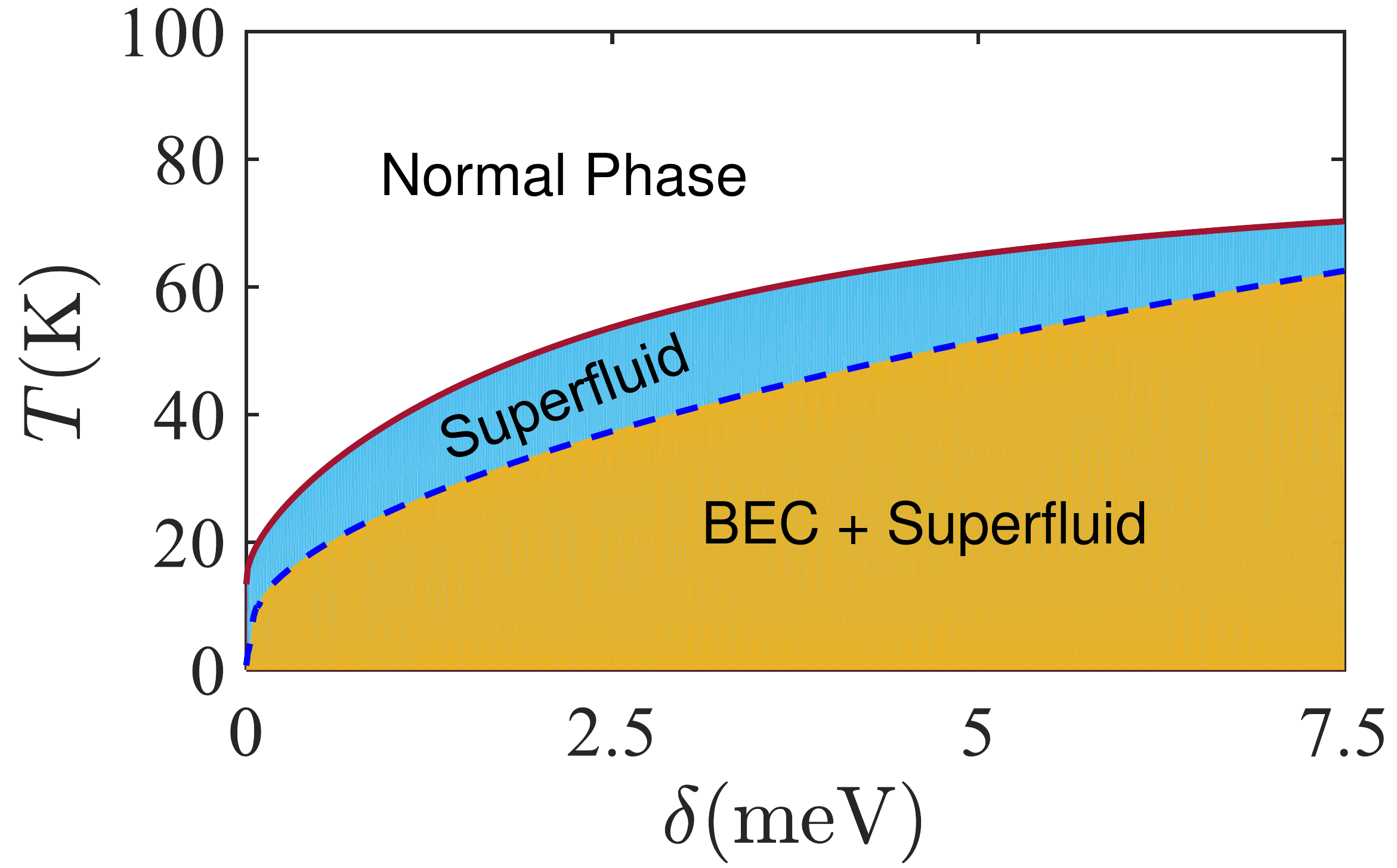}
\caption{Finite temperature phase diagram of intralayer excitons with change of $\delta$ and $T$. Three different phases
are shown in different color. The upper solid (lower dashed) line denotes the critical temperature $T_\mathrm{SF}$ ($T_\mathrm{BEC}$).
Exciton density $n=1.0*10^{11}$ cm$^{-2}$ is chosen here. 
\label{fig4}}
\end{figure}

We finally have enough information on the finite temperature phase diagram of this system in $\delta-T$ plane, as shown in Fig. \ref{fig4}. 
At $\delta = 0$, the BEC critical temperature is zero, while
the BKT transition temperature, given by the well-known Nelson-Kosterlitz relation \cite{nelson_Universal_1977}, is
nonzero. 
With the increase of $\delta$, the BEC critical temperature increases
and the superfluid critical temperature should also
increases, since both free vortices and non-singular excitations are asymptotically suppressed. For large $\delta$,
where the free vortices are completely suppressed for a given system size,
the superfluid density should continuously drop to zero,
similar to 3D case, where the superfluid critical temperature coincides with BEC
critical temperature. So both $T_{\mathrm{BEC}}$ and $T_{\mathrm{SF}}$ increase with $\delta$, and asymptotically approach each other.
One immediately realizes that the phase diagram consists of three different phases, i.e., BEC phase with long-range order (also a superfluid),
superfluid phase with quasi-long-range order, and a trivial normal phase.
In other words, at finite $\delta$, with the increase of temperature, the
system undergoes a two-step phase transition, first from
BEC to a non-BEC superfluid, and then to a normal phase.
Note that the superfluid critical temperature is only qualitatively demonstrated in Fig. \ref{fig4},
by interpolation between two known limiting cases ($\delta=0$ and large $\delta$). A quantitative treatment of superfluid phase transition,
taking full account of vortex-antivortex pair excitation induced screening,
using methods such as Monte Carlo simulation \cite{prokofev_Worm_2001,prokofev_Twodimensional_2002} will be postponed to a future work.

{\textit{Experimental observation}}. We now comment on the possible experimental realization of
 exciton {superfluid} in monolayer TMD. 
The major obstacle is the short exciton lifetime, whose magnitude relative to exciton thermalization time determines
whether thermal equilibrium can be reached. Recently, it is reported experimentally that an exciton
fluid at thermal equilibrium in monolayer MoS$_2$ was observed \cite{delaguila_Ultrafast_2023}.
This important finding points to the exciting possibilities for studying the exciton superfluid
in monolayer system and also demonstrates the experimental relevance of our theoretical model.
Using an experimental setup similar to Ref. \cite{sun_Enhanced_2022}, our predictions can be readily verified based on existing techniques.
The three different phases predicted in this work can be distinguished by measuring the spatial and temporal coherence of emitted photons,
from which we can infer the decay behaviour of exciton correlation function.
Similar techniques have been used in previous experimental demonstration of BKT transition in exciton-polariton condensate \cite{caputo_Topological_2018} 
and dipolar excitons \cite{dang_Observation_2020}. 
Note that in the presence of electron/hole doping, the screened Coulomb interaction is better treated by the Thomas-Fermi screening, which will turn the single exciton dispersion from linear to the conventional parabolic type \cite{Supplement_Material}, and hence should be avoided in experiments to observe the phenomena predicted here.

In summary, we have studied the unusual properties of exciton superfluid in monolayer TMD, 
possibly realizable based on recent experimental progress \cite{delaguila_Ultrafast_2023}, which is an unique bosonic system beyond conventional paradigm. 
With the valley-orbit coupling and valley Zeeman energy,
one can manipulate the dispersion of exciton center-of-mass motion. The attainable linear dispersion in lower
branch brings a variety of exotic properties of excitons associated with BEC and superfluidity. The BEC critical temperature is
nonzero, thereby
realizing a true condensate with long-range order in 2D homogeneous space. 
The form of vortex energy has an additional linear term, giving rise to superfluid transition different with BKT type.
There is a two-step phase transition with decrease of temperature from a normal exciton phase: at $T_{\mathrm{c1}}=T_{\mathrm{SF}}$ the system first enters
a superfluid phase with quasi-long range order, and then at $T_{\mathrm{c2}}=T_{\mathrm{BEC}}$ it enters the BEC phase with both long-range order
and superfluid properties. These interesting features can be possibly verified with the recent experimental progress on the exciton fluid in this system, by measuring spatial and temporal coherence of emitted photons.

{\textit{Acknowledgements.}} We are grateful to Yi-Cai Zhang for valuable discussions.
W.C. and Q.Z. are supported by NKRDPC (Grant No. 2022YFA1405304), NSFC (Grant No. 12004118), and Guangdong Basic and
Applied Basic Research Foundation (Grants No. 2020A1515110228 and No.
2021A1515010212). C.-J.H. gratefully acknowledges research support from
Gang Chen by the Research Grants Council of Hong
Kong with General Research Fund Grant No. 17306520.


\begin{thebibliography}{64}%
\makeatletter
\providecommand \@ifxundefined [1]{%
 \@ifx{#1\undefined}
}%
\providecommand \@ifnum [1]{%
 \ifnum #1\expandafter \@firstoftwo
 \else \expandafter \@secondoftwo
 \fi
}%
\providecommand \@ifx [1]{%
 \ifx #1\expandafter \@firstoftwo
 \else \expandafter \@secondoftwo
 \fi
}%
\providecommand \natexlab [1]{#1}%
\providecommand \enquote  [1]{``#1''}%
\providecommand \bibnamefont  [1]{#1}%
\providecommand \bibfnamefont [1]{#1}%
\providecommand \citenamefont [1]{#1}%
\providecommand \href@noop [0]{\@secondoftwo}%
\providecommand \href [0]{\begingroup \@sanitize@url \@href}%
\providecommand \@href[1]{\@@startlink{#1}\@@href}%
\providecommand \@@href[1]{\endgroup#1\@@endlink}%
\providecommand \@sanitize@url [0]{\catcode `\\12\catcode `\$12\catcode
  `\&12\catcode `\#12\catcode `\^12\catcode `\_12\catcode `\%12\relax}%
\providecommand \@@startlink[1]{}%
\providecommand \@@endlink[0]{}%
\providecommand \url  [0]{\begingroup\@sanitize@url \@url }%
\providecommand \@url [1]{\endgroup\@href {#1}{\urlprefix }}%
\providecommand \urlprefix  [0]{URL }%
\providecommand \Eprint [0]{\href }%
\providecommand \doibase [0]{https://doi.org/}%
\providecommand \selectlanguage [0]{\@gobble}%
\providecommand \bibinfo  [0]{\@secondoftwo}%
\providecommand \bibfield  [0]{\@secondoftwo}%
\providecommand \translation [1]{[#1]}%
\providecommand \BibitemOpen [0]{}%
\providecommand \bibitemStop [0]{}%
\providecommand \bibitemNoStop [0]{.\EOS\space}%
\providecommand \EOS [0]{\spacefactor3000\relax}%
\providecommand \BibitemShut  [1]{\csname bibitem#1\endcsname}%
\let\auto@bib@innerbib\@empty
\bibitem [{\citenamefont {Mermin}\ and\ \citenamefont
  {Wagner}(1966)}]{mermin_Absence_1966}%
  \BibitemOpen
  \bibfield  {author} {\bibinfo {author} {\bibfnamefont {N.~D.}\ \bibnamefont
  {Mermin}}\ and\ \bibinfo {author} {\bibfnamefont {H.}~\bibnamefont
  {Wagner}},\ }\href {https://doi.org/10.1103/PhysRevLett.17.1133} {\bibfield
  {journal} {\bibinfo  {journal} {Phys. Rev. Lett.}\ }\textbf {\bibinfo
  {volume} {17}},\ \bibinfo {pages} {1133} (\bibinfo {year}
  {1966})}\BibitemShut {NoStop}%
\bibitem [{\citenamefont {Hohenberg}(1967)}]{hohenberg_Existence_1967}%
  \BibitemOpen
  \bibfield  {author} {\bibinfo {author} {\bibfnamefont {P.~C.}\ \bibnamefont
  {Hohenberg}},\ }\href {https://doi.org/10.1103/PhysRev.158.383} {\bibfield
  {journal} {\bibinfo  {journal} {Phys. Rev.}\ }\textbf {\bibinfo {volume}
  {158}},\ \bibinfo {pages} {383} (\bibinfo {year} {1967})}\BibitemShut
  {NoStop}%
\bibitem [{\citenamefont {Berezinsky}(1972)}]{berezinsky_DESTRUCTION_1972}%
  \BibitemOpen
  \bibfield  {author} {\bibinfo {author} {\bibfnamefont {V.~L.}\ \bibnamefont
  {Berezinsky}},\ }\href@noop {} {\bibfield  {journal} {\bibinfo  {journal}
  {Sov. Phys. JETP}\ }\textbf {\bibinfo {volume} {34}},\ \bibinfo {pages} {610}
  (\bibinfo {year} {1972})}\BibitemShut {NoStop}%
\bibitem [{\citenamefont {Kosterlitz}\ and\ \citenamefont
  {Thouless}(1973)}]{kosterlitz_Ordering_1973}%
  \BibitemOpen
  \bibfield  {author} {\bibinfo {author} {\bibfnamefont {J.~M.}\ \bibnamefont
  {Kosterlitz}}\ and\ \bibinfo {author} {\bibfnamefont {D.~J.}\ \bibnamefont
  {Thouless}},\ }\href@noop {} {\bibfield  {journal} {\bibinfo  {journal} {J.
  Phys. C}\ }\textbf {\bibinfo {volume} {6}},\ \bibinfo {pages} {1181}
  (\bibinfo {year} {1973})}\BibitemShut {NoStop}%
\bibitem [{\citenamefont {Bishop}\ and\ \citenamefont
  {Reppy}(1978)}]{bishop_Study_1978}%
  \BibitemOpen
  \bibfield  {author} {\bibinfo {author} {\bibfnamefont {D.~J.}\ \bibnamefont
  {Bishop}}\ and\ \bibinfo {author} {\bibfnamefont {J.~D.}\ \bibnamefont
  {Reppy}},\ }\href {https://doi.org/10.1103/PhysRevLett.40.1727} {\bibfield
  {journal} {\bibinfo  {journal} {Phys. Rev. Lett.}\ }\textbf {\bibinfo
  {volume} {40}},\ \bibinfo {pages} {1727} (\bibinfo {year}
  {1978})}\BibitemShut {NoStop}%
\bibitem [{\citenamefont {Resnick}\ \emph {et~al.}(1981)\citenamefont
  {Resnick}, \citenamefont {Garland}, \citenamefont {Boyd}, \citenamefont
  {Shoemaker},\ and\ \citenamefont
  {Newrock}}]{resnick_KosterlitzThouless_1981}%
  \BibitemOpen
  \bibfield  {author} {\bibinfo {author} {\bibfnamefont {D.~J.}\ \bibnamefont
  {Resnick}}, \bibinfo {author} {\bibfnamefont {J.~C.}\ \bibnamefont
  {Garland}}, \bibinfo {author} {\bibfnamefont {J.~T.}\ \bibnamefont {Boyd}},
  \bibinfo {author} {\bibfnamefont {S.}~\bibnamefont {Shoemaker}},\ and\
  \bibinfo {author} {\bibfnamefont {R.~S.}\ \bibnamefont {Newrock}},\ }\href
  {https://doi.org/10.1103/PhysRevLett.47.1542} {\bibfield  {journal} {\bibinfo
   {journal} {Phys. Rev. Lett.}\ }\textbf {\bibinfo {volume} {47}},\ \bibinfo
  {pages} {1542} (\bibinfo {year} {1981})}\BibitemShut {NoStop}%
\bibitem [{\citenamefont {Hadzibabic}\ \emph {et~al.}(2006)\citenamefont
  {Hadzibabic}, \citenamefont {Kr{\"u}ger}, \citenamefont {Cheneau},
  \citenamefont {Battelier},\ and\ \citenamefont
  {Dalibard}}]{hadzibabic_Berezinskii_2006}%
  \BibitemOpen
  \bibfield  {author} {\bibinfo {author} {\bibfnamefont {Z.}~\bibnamefont
  {Hadzibabic}}, \bibinfo {author} {\bibfnamefont {P.}~\bibnamefont
  {Kr{\"u}ger}}, \bibinfo {author} {\bibfnamefont {M.}~\bibnamefont {Cheneau}},
  \bibinfo {author} {\bibfnamefont {B.}~\bibnamefont {Battelier}},\ and\
  \bibinfo {author} {\bibfnamefont {J.}~\bibnamefont {Dalibard}},\ }\href
  {https://doi.org/10.1038/nature04851} {\bibfield  {journal} {\bibinfo
  {journal} {Nature}\ }\textbf {\bibinfo {volume} {441}},\ \bibinfo {pages}
  {1118} (\bibinfo {year} {2006})}\BibitemShut {NoStop}%
\bibitem [{\citenamefont {Murthy}\ \emph {et~al.}(2015)\citenamefont {Murthy},
  \citenamefont {Boettcher}, \citenamefont {Bayha}, \citenamefont {Holzmann},
  \citenamefont {Kedar}, \citenamefont {Neidig}, \citenamefont {Ries},
  \citenamefont {Wenz}, \citenamefont {Z{\"u}rn},\ and\ \citenamefont
  {Jochim}}]{murthy_Observation_2015}%
  \BibitemOpen
  \bibfield  {author} {\bibinfo {author} {\bibfnamefont {P.~A.}\ \bibnamefont
  {Murthy}}, \bibinfo {author} {\bibfnamefont {I.}~\bibnamefont {Boettcher}},
  \bibinfo {author} {\bibfnamefont {L.}~\bibnamefont {Bayha}}, \bibinfo
  {author} {\bibfnamefont {M.}~\bibnamefont {Holzmann}}, \bibinfo {author}
  {\bibfnamefont {D.}~\bibnamefont {Kedar}}, \bibinfo {author} {\bibfnamefont
  {M.}~\bibnamefont {Neidig}}, \bibinfo {author} {\bibfnamefont {M.~G.}\
  \bibnamefont {Ries}}, \bibinfo {author} {\bibfnamefont {A.~N.}\ \bibnamefont
  {Wenz}}, \bibinfo {author} {\bibfnamefont {G.}~\bibnamefont {Z{\"u}rn}},\
  and\ \bibinfo {author} {\bibfnamefont {S.}~\bibnamefont {Jochim}},\ }\href
  {https://doi.org/10.1103/PhysRevLett.115.010401} {\bibfield  {journal}
  {\bibinfo  {journal} {Phys. Rev. Lett.}\ }\textbf {\bibinfo {volume} {115}},\
  \bibinfo {pages} {010401} (\bibinfo {year} {2015})}\BibitemShut {NoStop}%
\bibitem [{\citenamefont {Caputo}\ \emph {et~al.}(2018)\citenamefont {Caputo},
  \citenamefont {Ballarini}, \citenamefont {Dagvadorj}, \citenamefont
  {S{\'a}nchez~Mu{\~n}oz}, \citenamefont {De~Giorgi}, \citenamefont {Dominici},
  \citenamefont {West}, \citenamefont {Pfeiffer}, \citenamefont {Gigli},
  \citenamefont {Laussy}, \citenamefont {Szyma{\'n}ska},\ and\ \citenamefont
  {Sanvitto}}]{caputo_Topological_2018}%
  \BibitemOpen
  \bibfield  {author} {\bibinfo {author} {\bibfnamefont {D.}~\bibnamefont
  {Caputo}}, \bibinfo {author} {\bibfnamefont {D.}~\bibnamefont {Ballarini}},
  \bibinfo {author} {\bibfnamefont {G.}~\bibnamefont {Dagvadorj}}, \bibinfo
  {author} {\bibfnamefont {C.}~\bibnamefont {S{\'a}nchez~Mu{\~n}oz}}, \bibinfo
  {author} {\bibfnamefont {M.}~\bibnamefont {De~Giorgi}}, \bibinfo {author}
  {\bibfnamefont {L.}~\bibnamefont {Dominici}}, \bibinfo {author}
  {\bibfnamefont {K.}~\bibnamefont {West}}, \bibinfo {author} {\bibfnamefont
  {L.~N.}\ \bibnamefont {Pfeiffer}}, \bibinfo {author} {\bibfnamefont
  {G.}~\bibnamefont {Gigli}}, \bibinfo {author} {\bibfnamefont {F.~P.}\
  \bibnamefont {Laussy}}, \bibinfo {author} {\bibfnamefont {M.~H.}\
  \bibnamefont {Szyma{\'n}ska}},\ and\ \bibinfo {author} {\bibfnamefont
  {D.}~\bibnamefont {Sanvitto}},\ }\href {https://doi.org/10.1038/nmat5039}
  {\bibfield  {journal} {\bibinfo  {journal} {Nat. Mater.}\ }\textbf {\bibinfo
  {volume} {17}},\ \bibinfo {pages} {145} (\bibinfo {year} {2018})}\BibitemShut
  {NoStop}%
\bibitem [{\citenamefont {Dang}\ \emph {et~al.}(2020)\citenamefont {Dang},
  \citenamefont {Zamorano}, \citenamefont {Suffit}, \citenamefont {West},
  \citenamefont {Baldwin}, \citenamefont {Pfeiffer}, \citenamefont {Holzmann},\
  and\ \citenamefont {Dubin}}]{dang_Observation_2020}%
  \BibitemOpen
  \bibfield  {author} {\bibinfo {author} {\bibfnamefont {S.}~\bibnamefont
  {Dang}}, \bibinfo {author} {\bibfnamefont {M.}~\bibnamefont {Zamorano}},
  \bibinfo {author} {\bibfnamefont {S.}~\bibnamefont {Suffit}}, \bibinfo
  {author} {\bibfnamefont {K.}~\bibnamefont {West}}, \bibinfo {author}
  {\bibfnamefont {K.}~\bibnamefont {Baldwin}}, \bibinfo {author} {\bibfnamefont
  {L.}~\bibnamefont {Pfeiffer}}, \bibinfo {author} {\bibfnamefont
  {M.}~\bibnamefont {Holzmann}},\ and\ \bibinfo {author} {\bibfnamefont
  {F.}~\bibnamefont {Dubin}},\ }\href
  {https://doi.org/10.1103/PhysRevResearch.2.032013} {\bibfield  {journal}
  {\bibinfo  {journal} {Phys. Rev. Res.}\ }\textbf {\bibinfo {volume} {2}},\
  \bibinfo {pages} {032013(R)} (\bibinfo {year} {2020})}\BibitemShut {NoStop}%
\bibitem [{\citenamefont {Dalfovo}\ \emph {et~al.}(1999)\citenamefont
  {Dalfovo}, \citenamefont {Giorgini}, \citenamefont {Pitaevskii},\ and\
  \citenamefont {Stringari}}]{dalfovo_Theory_1999}%
  \BibitemOpen
  \bibfield  {author} {\bibinfo {author} {\bibfnamefont {F.}~\bibnamefont
  {Dalfovo}}, \bibinfo {author} {\bibfnamefont {S.}~\bibnamefont {Giorgini}},
  \bibinfo {author} {\bibfnamefont {L.~P.}\ \bibnamefont {Pitaevskii}},\ and\
  \bibinfo {author} {\bibfnamefont {S.}~\bibnamefont {Stringari}},\ }\href
  {https://doi.org/10.1103/RevModPhys.71.463} {\bibfield  {journal} {\bibinfo
  {journal} {Rev. Mod. Phys.}\ }\textbf {\bibinfo {volume} {71}},\ \bibinfo
  {pages} {463} (\bibinfo {year} {1999})}\BibitemShut {NoStop}%
\bibitem [{\citenamefont {Tononi}\ and\ \citenamefont
  {Salasnich}(2019)}]{tononi_BoseEinstein_2019}%
  \BibitemOpen
  \bibfield  {author} {\bibinfo {author} {\bibfnamefont {A.}~\bibnamefont
  {Tononi}}\ and\ \bibinfo {author} {\bibfnamefont {L.}~\bibnamefont
  {Salasnich}},\ }\href {https://doi.org/10.1103/PhysRevLett.123.160403}
  {\bibfield  {journal} {\bibinfo  {journal} {Phys. Rev. Lett.}\ }\textbf
  {\bibinfo {volume} {123}},\ \bibinfo {pages} {160403} (\bibinfo {year}
  {2019})}\BibitemShut {NoStop}%
\bibitem [{\citenamefont {Po}\ and\ \citenamefont
  {Zhou}(2015)}]{po_Twodimensional_2015}%
  \BibitemOpen
  \bibfield  {author} {\bibinfo {author} {\bibfnamefont {H.~C.}\ \bibnamefont
  {Po}}\ and\ \bibinfo {author} {\bibfnamefont {Q.}~\bibnamefont {Zhou}},\
  }\href {https://doi.org/10.1038/ncomms9012} {\bibfield  {journal} {\bibinfo
  {journal} {Nat. Commun.}\ }\textbf {\bibinfo {volume} {6}},\ \bibinfo {pages}
  {8012} (\bibinfo {year} {2015})}\BibitemShut {NoStop}%
\bibitem [{\citenamefont {Berman}\ \emph {et~al.}(2012)\citenamefont {Berman},
  \citenamefont {Kezerashvili},\ and\ \citenamefont
  {Ziegler}}]{berman_Superfluidity_2012}%
  \BibitemOpen
  \bibfield  {author} {\bibinfo {author} {\bibfnamefont {O.~L.}\ \bibnamefont
  {Berman}}, \bibinfo {author} {\bibfnamefont {R.~Y.}\ \bibnamefont
  {Kezerashvili}},\ and\ \bibinfo {author} {\bibfnamefont {K.}~\bibnamefont
  {Ziegler}},\ }\href {https://doi.org/10.1103/PhysRevB.85.035418} {\bibfield
  {journal} {\bibinfo  {journal} {Phys. Rev. B}\ }\textbf {\bibinfo {volume}
  {85}},\ \bibinfo {pages} {035418} (\bibinfo {year} {2012})}\BibitemShut
  {NoStop}%
\bibitem [{\citenamefont {Fogler}\ \emph {et~al.}(2014)\citenamefont {Fogler},
  \citenamefont {Butov},\ and\ \citenamefont
  {Novoselov}}]{fogler_Hightemperature_2014}%
  \BibitemOpen
  \bibfield  {author} {\bibinfo {author} {\bibfnamefont {M.~M.}\ \bibnamefont
  {Fogler}}, \bibinfo {author} {\bibfnamefont {L.~V.}\ \bibnamefont {Butov}},\
  and\ \bibinfo {author} {\bibfnamefont {K.~S.}\ \bibnamefont {Novoselov}},\
  }\href {https://doi.org/10.1038/ncomms5555} {\bibfield  {journal} {\bibinfo
  {journal} {Nat. Commun.}\ }\textbf {\bibinfo {volume} {5}},\ \bibinfo {pages}
  {4555} (\bibinfo {year} {2014})}\BibitemShut {NoStop}%
\bibitem [{\citenamefont {Wu}\ \emph {et~al.}(2015{\natexlab{a}})\citenamefont
  {Wu}, \citenamefont {Xue},\ and\ \citenamefont {MacDonald}}]{wu_Theory_2015}%
  \BibitemOpen
  \bibfield  {author} {\bibinfo {author} {\bibfnamefont {F.-C.}\ \bibnamefont
  {Wu}}, \bibinfo {author} {\bibfnamefont {F.}~\bibnamefont {Xue}},\ and\
  \bibinfo {author} {\bibfnamefont {A.~H.}\ \bibnamefont {MacDonald}},\ }\href
  {https://doi.org/10.1103/PhysRevB.92.165121} {\bibfield  {journal} {\bibinfo
  {journal} {Phys. Rev. B}\ }\textbf {\bibinfo {volume} {92}},\ \bibinfo
  {pages} {165121} (\bibinfo {year} {2015}{\natexlab{a}})}\BibitemShut
  {NoStop}%
\bibitem [{\citenamefont {Berman}\ and\ \citenamefont
  {Kezerashvili}(2016)}]{berman_Hightemperature_2016}%
  \BibitemOpen
  \bibfield  {author} {\bibinfo {author} {\bibfnamefont {O.~L.}\ \bibnamefont
  {Berman}}\ and\ \bibinfo {author} {\bibfnamefont {R.~Y.}\ \bibnamefont
  {Kezerashvili}},\ }\href {https://doi.org/10.1103/PhysRevB.93.245410}
  {\bibfield  {journal} {\bibinfo  {journal} {Phys. Rev. B}\ }\textbf {\bibinfo
  {volume} {93}},\ \bibinfo {pages} {245410} (\bibinfo {year}
  {2016})}\BibitemShut {NoStop}%
\bibitem [{\citenamefont {Berman}\ and\ \citenamefont
  {Kezerashvili}(2017)}]{berman_Superfluidity_2017}%
  \BibitemOpen
  \bibfield  {author} {\bibinfo {author} {\bibfnamefont {O.~L.}\ \bibnamefont
  {Berman}}\ and\ \bibinfo {author} {\bibfnamefont {R.~Y.}\ \bibnamefont
  {Kezerashvili}},\ }\href {https://doi.org/10.1103/PhysRevB.96.094502}
  {\bibfield  {journal} {\bibinfo  {journal} {Phys. Rev. B}\ }\textbf {\bibinfo
  {volume} {96}},\ \bibinfo {pages} {094502} (\bibinfo {year}
  {2017})}\BibitemShut {NoStop}%
\bibitem [{\citenamefont {Debnath}\ \emph {et~al.}(2017)\citenamefont
  {Debnath}, \citenamefont {Barlas}, \citenamefont {Wickramaratne},
  \citenamefont {Neupane},\ and\ \citenamefont {Lake}}]{debnath_Exciton_2017}%
  \BibitemOpen
  \bibfield  {author} {\bibinfo {author} {\bibfnamefont {B.}~\bibnamefont
  {Debnath}}, \bibinfo {author} {\bibfnamefont {Y.}~\bibnamefont {Barlas}},
  \bibinfo {author} {\bibfnamefont {D.}~\bibnamefont {Wickramaratne}}, \bibinfo
  {author} {\bibfnamefont {M.~R.}\ \bibnamefont {Neupane}},\ and\ \bibinfo
  {author} {\bibfnamefont {R.~K.}\ \bibnamefont {Lake}},\ }\href
  {https://doi.org/10.1103/PhysRevB.96.174504} {\bibfield  {journal} {\bibinfo
  {journal} {Phys. Rev. B}\ }\textbf {\bibinfo {volume} {96}},\ \bibinfo
  {pages} {174504} (\bibinfo {year} {2017})}\BibitemShut {NoStop}%
\bibitem [{\citenamefont {Zhu}\ \emph {et~al.}(2019)\citenamefont {Zhu},
  \citenamefont {Tu}, \citenamefont {Tong},\ and\ \citenamefont
  {Yao}}]{zhu_Gate_2019}%
  \BibitemOpen
  \bibfield  {author} {\bibinfo {author} {\bibfnamefont {Q.}~\bibnamefont
  {Zhu}}, \bibinfo {author} {\bibfnamefont {M.~W.-Y.}\ \bibnamefont {Tu}},
  \bibinfo {author} {\bibfnamefont {Q.}~\bibnamefont {Tong}},\ and\ \bibinfo
  {author} {\bibfnamefont {W.}~\bibnamefont {Yao}},\ }\href
  {https://doi.org/10.1126/sciadv.aau6120} {\bibfield  {journal} {\bibinfo
  {journal} {Sci. Adv.}\ }\textbf {\bibinfo {volume} {5}},\ \bibinfo {pages}
  {eaau6120} (\bibinfo {year} {2019})}\BibitemShut {NoStop}%
\bibitem [{\citenamefont {Wang}\ \emph {et~al.}(2019)\citenamefont {Wang},
  \citenamefont {Rhodes}, \citenamefont {Watanabe}, \citenamefont {Taniguchi},
  \citenamefont {Hone}, \citenamefont {Shan},\ and\ \citenamefont
  {Mak}}]{wang_Evidence_2019}%
  \BibitemOpen
  \bibfield  {author} {\bibinfo {author} {\bibfnamefont {Z.}~\bibnamefont
  {Wang}}, \bibinfo {author} {\bibfnamefont {D.~A.}\ \bibnamefont {Rhodes}},
  \bibinfo {author} {\bibfnamefont {K.}~\bibnamefont {Watanabe}}, \bibinfo
  {author} {\bibfnamefont {T.}~\bibnamefont {Taniguchi}}, \bibinfo {author}
  {\bibfnamefont {J.~C.}\ \bibnamefont {Hone}}, \bibinfo {author}
  {\bibfnamefont {J.}~\bibnamefont {Shan}},\ and\ \bibinfo {author}
  {\bibfnamefont {K.~F.}\ \bibnamefont {Mak}},\ }\href
  {https://doi.org/10.1038/s41586-019-1591-7} {\bibfield  {journal} {\bibinfo
  {journal} {Nature}\ }\textbf {\bibinfo {volume} {574}},\ \bibinfo {pages}
  {76} (\bibinfo {year} {2019})}\BibitemShut {NoStop}%
\bibitem [{\citenamefont {Ma}\ \emph {et~al.}(2021)\citenamefont {Ma},
  \citenamefont {Nguyen}, \citenamefont {Wang}, \citenamefont {Zeng},
  \citenamefont {Watanabe}, \citenamefont {Taniguchi}, \citenamefont
  {MacDonald}, \citenamefont {Mak},\ and\ \citenamefont
  {Shan}}]{ma_Strongly_2021}%
  \BibitemOpen
  \bibfield  {author} {\bibinfo {author} {\bibfnamefont {L.}~\bibnamefont
  {Ma}}, \bibinfo {author} {\bibfnamefont {P.~X.}\ \bibnamefont {Nguyen}},
  \bibinfo {author} {\bibfnamefont {Z.}~\bibnamefont {Wang}}, \bibinfo {author}
  {\bibfnamefont {Y.}~\bibnamefont {Zeng}}, \bibinfo {author} {\bibfnamefont
  {K.}~\bibnamefont {Watanabe}}, \bibinfo {author} {\bibfnamefont
  {T.}~\bibnamefont {Taniguchi}}, \bibinfo {author} {\bibfnamefont {A.~H.}\
  \bibnamefont {MacDonald}}, \bibinfo {author} {\bibfnamefont {K.~F.}\
  \bibnamefont {Mak}},\ and\ \bibinfo {author} {\bibfnamefont {J.}~\bibnamefont
  {Shan}},\ }\href {https://doi.org/10.1038/s41586-021-03947-9} {\bibfield
  {journal} {\bibinfo  {journal} {Nature}\ }\textbf {\bibinfo {volume} {598}},\
  \bibinfo {pages} {585} (\bibinfo {year} {2021})}\BibitemShut {NoStop}%
\bibitem [{\citenamefont {Remez}\ and\ \citenamefont
  {Cooper}(2022)}]{remez_Leaky_2022}%
  \BibitemOpen
  \bibfield  {author} {\bibinfo {author} {\bibfnamefont {B.}~\bibnamefont
  {Remez}}\ and\ \bibinfo {author} {\bibfnamefont {N.~R.}\ \bibnamefont
  {Cooper}},\ }\href {https://doi.org/10.1103/PhysRevResearch.4.L022042}
  {\bibfield  {journal} {\bibinfo  {journal} {Phys. Rev. Res.}\ }\textbf
  {\bibinfo {volume} {4}},\ \bibinfo {pages} {L022042} (\bibinfo {year}
  {2022})}\BibitemShut {NoStop}%
\bibitem [{\citenamefont {Deng}\ \emph {et~al.}(2022)\citenamefont {Deng},
  \citenamefont {Chu},\ and\ \citenamefont {Zhu}}]{deng_Moir_2022}%
  \BibitemOpen
  \bibfield  {author} {\bibinfo {author} {\bibfnamefont {S.}~\bibnamefont
  {Deng}}, \bibinfo {author} {\bibfnamefont {Y.}~\bibnamefont {Chu}},\ and\
  \bibinfo {author} {\bibfnamefont {Q.}~\bibnamefont {Zhu}},\ }\href
  {https://doi.org/10.1103/PhysRevB.106.155410} {\bibfield  {journal} {\bibinfo
   {journal} {Phys. Rev. B}\ }\textbf {\bibinfo {volume} {106}},\ \bibinfo
  {pages} {155410} (\bibinfo {year} {2022})}\BibitemShut {NoStop}%
\bibitem [{\citenamefont {Zimmerman}\ \emph {et~al.}(2022)\citenamefont
  {Zimmerman}, \citenamefont {Rapaport},\ and\ \citenamefont
  {Gazit}}]{zimmerman_Collective_2022}%
  \BibitemOpen
  \bibfield  {author} {\bibinfo {author} {\bibfnamefont {M.}~\bibnamefont
  {Zimmerman}}, \bibinfo {author} {\bibfnamefont {R.}~\bibnamefont
  {Rapaport}},\ and\ \bibinfo {author} {\bibfnamefont {S.}~\bibnamefont
  {Gazit}},\ }\href {https://doi.org/10.1073/pnas.2205845119} {\bibfield
  {journal} {\bibinfo  {journal} {Proc. Natl. Acad. Sci. U.S.A.}\ }\textbf
  {\bibinfo {volume} {119}},\ \bibinfo {pages} {e2205845119} (\bibinfo {year}
  {2022})}\BibitemShut {NoStop}%
\bibitem [{\citenamefont {Guo}\ \emph {et~al.}(2022)\citenamefont {Guo},
  \citenamefont {Zhang},\ and\ \citenamefont {Lu}}]{guo_Tuning_2022}%
  \BibitemOpen
  \bibfield  {author} {\bibinfo {author} {\bibfnamefont {H.}~\bibnamefont
  {Guo}}, \bibinfo {author} {\bibfnamefont {X.}~\bibnamefont {Zhang}},\ and\
  \bibinfo {author} {\bibfnamefont {G.}~\bibnamefont {Lu}},\ }\href
  {https://doi.org/10.1126/sciadv.abp9757} {\bibfield  {journal} {\bibinfo
  {journal} {Sci. Adv.}\ }\textbf {\bibinfo {volume} {8}},\ \bibinfo {pages}
  {eabp9757} (\bibinfo {year} {2022})}\BibitemShut {NoStop}%
\bibitem [{\citenamefont {{del {\'A}guila}}\ \emph {et~al.}(2023)\citenamefont
  {{del {\'A}guila}}, \citenamefont {Wong}, \citenamefont {Wadgaonkar},
  \citenamefont {Fieramosca}, \citenamefont {Liu}, \citenamefont {Vaklinova},
  \citenamefont {Dal~Forno}, \citenamefont {Do}, \citenamefont {Wei},
  \citenamefont {Watanabe}, \citenamefont {Taniguchi}, \citenamefont
  {Novoselov}, \citenamefont {Koperski}, \citenamefont {Battiato},\ and\
  \citenamefont {Xiong}}]{delaguila_Ultrafast_2023}%
  \BibitemOpen
  \bibfield  {author} {\bibinfo {author} {\bibfnamefont {A.~G.}\ \bibnamefont
  {{del {\'A}guila}}}, \bibinfo {author} {\bibfnamefont {Y.~R.}\ \bibnamefont
  {Wong}}, \bibinfo {author} {\bibfnamefont {I.}~\bibnamefont {Wadgaonkar}},
  \bibinfo {author} {\bibfnamefont {A.}~\bibnamefont {Fieramosca}}, \bibinfo
  {author} {\bibfnamefont {X.}~\bibnamefont {Liu}}, \bibinfo {author}
  {\bibfnamefont {K.}~\bibnamefont {Vaklinova}}, \bibinfo {author}
  {\bibfnamefont {S.}~\bibnamefont {Dal~Forno}}, \bibinfo {author}
  {\bibfnamefont {T.~T.~H.}\ \bibnamefont {Do}}, \bibinfo {author}
  {\bibfnamefont {H.~Y.}\ \bibnamefont {Wei}}, \bibinfo {author} {\bibfnamefont
  {K.}~\bibnamefont {Watanabe}}, \bibinfo {author} {\bibfnamefont
  {T.}~\bibnamefont {Taniguchi}}, \bibinfo {author} {\bibfnamefont {K.~S.}\
  \bibnamefont {Novoselov}}, \bibinfo {author} {\bibfnamefont {M.}~\bibnamefont
  {Koperski}}, \bibinfo {author} {\bibfnamefont {M.}~\bibnamefont {Battiato}},\
  and\ \bibinfo {author} {\bibfnamefont {Q.}~\bibnamefont {Xiong}},\ }\href
  {https://doi.org/DOI:10.1038/s41565-023-01438-8} \bibfield
  {} {\bibinfo  {journal} {Nat. Nanotechnol.}\ }\textbf {\bibinfo {volume} {18}},\ \bibinfo {pages}{1012} (\bibinfo {year} {2023})\BibitemShut
  {NoStop}%
\bibitem [{\citenamefont {Yu}\ \emph {et~al.}(2014)\citenamefont {Yu},
  \citenamefont {Liu}, \citenamefont {Gong}, \citenamefont {Xu},\ and\
  \citenamefont {Yao}}]{yu_Dirac_2014}%
  \BibitemOpen
  \bibfield  {author} {\bibinfo {author} {\bibfnamefont {H.}~\bibnamefont
  {Yu}}, \bibinfo {author} {\bibfnamefont {G.-B.}\ \bibnamefont {Liu}},
  \bibinfo {author} {\bibfnamefont {P.}~\bibnamefont {Gong}}, \bibinfo {author}
  {\bibfnamefont {X.}~\bibnamefont {Xu}},\ and\ \bibinfo {author}
  {\bibfnamefont {W.}~\bibnamefont {Yao}},\ }\href
  {https://doi.org/10.1038/ncomms4876} {\bibfield  {journal} {\bibinfo
  {journal} {Nat. Commun.}\ }\textbf {\bibinfo {volume} {5}},\ \bibinfo {pages}
  {3876} (\bibinfo {year} {2014})}\BibitemShut {NoStop}%
\bibitem [{\citenamefont {Qiu}\ \emph {et~al.}(2015)\citenamefont {Qiu},
  \citenamefont {Cao},\ and\ \citenamefont {Louie}}]{qiu_Nonanalyticity_2015}%
  \BibitemOpen
  \bibfield  {author} {\bibinfo {author} {\bibfnamefont {D.~Y.}\ \bibnamefont
  {Qiu}}, \bibinfo {author} {\bibfnamefont {T.}~\bibnamefont {Cao}},\ and\
  \bibinfo {author} {\bibfnamefont {S.~G.}\ \bibnamefont {Louie}},\ }\href
  {https://doi.org/10.1103/PhysRevLett.115.176801} {\bibfield  {journal}
  {\bibinfo  {journal} {Phys. Rev. Lett.}\ }\textbf {\bibinfo {volume} {115}},\
  \bibinfo {pages} {176801} (\bibinfo {year} {2015})}\BibitemShut {NoStop}%
\bibitem [{\citenamefont {Berman}\ \emph {et~al.}(2017)\citenamefont {Berman},
  \citenamefont {Gumbs},\ and\ \citenamefont
  {Kezerashvili}}]{berman_BoseEinstein_2017}%
  \BibitemOpen
  \bibfield  {author} {\bibinfo {author} {\bibfnamefont {O.~L.}\ \bibnamefont
  {Berman}}, \bibinfo {author} {\bibfnamefont {G.}~\bibnamefont {Gumbs}},\ and\
  \bibinfo {author} {\bibfnamefont {R.~Y.}\ \bibnamefont {Kezerashvili}},\
  }\href {https://doi.org/10.1103/PhysRevB.96.014505} {\bibfield  {journal}
  {\bibinfo  {journal} {Phys. Rev. B}\ }\textbf {\bibinfo {volume} {96}},\
  \bibinfo {pages} {014505} (\bibinfo {year} {2017})}\BibitemShut {NoStop}%
\bibitem [{\citenamefont {{Saberi-Pouya}}\ \emph {et~al.}(2018)\citenamefont
  {{Saberi-Pouya}}, \citenamefont {Zarenia}, \citenamefont {Perali},
  \citenamefont {Vazifehshenas},\ and\ \citenamefont
  {Peeters}}]{saberi-pouya_Hightemperature_2018}%
  \BibitemOpen
  \bibfield  {author} {\bibinfo {author} {\bibfnamefont {S.}~\bibnamefont
  {{Saberi-Pouya}}}, \bibinfo {author} {\bibfnamefont {M.}~\bibnamefont
  {Zarenia}}, \bibinfo {author} {\bibfnamefont {A.}~\bibnamefont {Perali}},
  \bibinfo {author} {\bibfnamefont {T.}~\bibnamefont {Vazifehshenas}},\ and\
  \bibinfo {author} {\bibfnamefont {F.~M.}\ \bibnamefont {Peeters}},\ }\href
  {https://doi.org/10.1103/PhysRevB.97.174503} {\bibfield  {journal} {\bibinfo
  {journal} {Phys. Rev. B}\ }\textbf {\bibinfo {volume} {97}},\ \bibinfo
  {pages} {174503} (\bibinfo {year} {2018})}\BibitemShut {NoStop}%
\bibitem [{\citenamefont {Kezerashvili}\ and\ \citenamefont
  {Spiridonova}(2022)}]{kezerashvili_Superfluidity_2022}%
  \BibitemOpen
  \bibfield  {author} {\bibinfo {author} {\bibfnamefont {R.~Y.}\ \bibnamefont
  {Kezerashvili}}\ and\ \bibinfo {author} {\bibfnamefont {A.}~\bibnamefont
  {Spiridonova}},\ }\href {https://doi.org/10.1103/PhysRevB.106.245306}
  {\bibfield  {journal} {\bibinfo  {journal} {Phys. Rev. B}\ }\textbf {\bibinfo
  {volume} {106}},\ \bibinfo {pages} {245306} (\bibinfo {year}
  {2022})}\BibitemShut {NoStop}%
\bibitem [{\citenamefont {Wu}\ \emph {et~al.}(2015{\natexlab{c}})\citenamefont
  {Wu}, \citenamefont {Qu},\ and\ \citenamefont {MacDonald}}]{wu_Exciton_2015}%
  \BibitemOpen
  \bibfield  {author} {\bibinfo {author} {\bibfnamefont {F.}~\bibnamefont
  {Wu}}, \bibinfo {author} {\bibfnamefont {F.}~\bibnamefont {Qu}},\ and\
  \bibinfo {author} {\bibfnamefont {A.~H.}\ \bibnamefont {MacDonald}},\ }\href
  {https://doi.org/10.1103/PhysRevB.91.075310} {\bibfield  {journal} {\bibinfo
  {journal} {Phys. Rev. B}\ }\textbf {\bibinfo {volume} {91}},\ \bibinfo
  {pages} {075310} (\bibinfo {year} {2015}{\natexlab{c}})}\BibitemShut
  {NoStop}%
\bibitem [{\citenamefont {Wu}\ \emph {et~al.}(2017)\citenamefont {Wu},
  \citenamefont {Lovorn},\ and\ \citenamefont
  {MacDonald}}]{wu_Topological_2017}%
  \BibitemOpen
  \bibfield  {author} {\bibinfo {author} {\bibfnamefont {F.}~\bibnamefont
  {Wu}}, \bibinfo {author} {\bibfnamefont {T.}~\bibnamefont {Lovorn}},\ and\
  \bibinfo {author} {\bibfnamefont {A.~H.}\ \bibnamefont {MacDonald}},\ }\href
  {https://doi.org/10.1103/PhysRevLett.118.147401} {\bibfield  {journal}
  {\bibinfo  {journal} {Phys. Rev. Lett.}\ }\textbf {\bibinfo {volume} {118}},\
  \bibinfo {pages} {147401} (\bibinfo {year} {2017})}\BibitemShut {NoStop}%
\bibitem [{\citenamefont {Sauer}\ \emph {et~al.}(2021)\citenamefont {Sauer},
  \citenamefont {Nielsen}, \citenamefont {{Merring-Mikkelsen}},\ and\
  \citenamefont {Pedersen}}]{sauer_Optical_2021}%
  \BibitemOpen
  \bibfield  {author} {\bibinfo {author} {\bibfnamefont {M.~O.}\ \bibnamefont
  {Sauer}}, \bibinfo {author} {\bibfnamefont {C.~E.~M.}\ \bibnamefont
  {Nielsen}}, \bibinfo {author} {\bibfnamefont {L.}~\bibnamefont
  {{Merring-Mikkelsen}}},\ and\ \bibinfo {author} {\bibfnamefont {T.~G.}\
  \bibnamefont {Pedersen}},\ }\href
  {https://doi.org/10.1103/PhysRevB.103.205404} {\bibfield  {journal} {\bibinfo
   {journal} {Phys. Rev. B}\ }\textbf {\bibinfo {volume} {103}},\ \bibinfo
  {pages} {205404} (\bibinfo {year} {2021})}\BibitemShut {NoStop}%
\bibitem [{\citenamefont {Salvador}\ \emph {et~al.}(2022)\citenamefont
  {Salvador}, \citenamefont {Kuhlenkamp}, \citenamefont {Ciorciaro},
  \citenamefont {Knap},\ and\ \citenamefont {{\.I}mamo{\u
  g}lu}}]{salvador_Optical_2022}%
  \BibitemOpen
  \bibfield  {author} {\bibinfo {author} {\bibfnamefont {A.~G.}\ \bibnamefont
  {Salvador}}, \bibinfo {author} {\bibfnamefont {C.}~\bibnamefont
  {Kuhlenkamp}}, \bibinfo {author} {\bibfnamefont {L.}~\bibnamefont
  {Ciorciaro}}, \bibinfo {author} {\bibfnamefont {M.}~\bibnamefont {Knap}},\
  and\ \bibinfo {author} {\bibfnamefont {A.}~\bibnamefont {{\.I}mamo{\u
  g}lu}},\ }\href {https://doi.org/10.1103/PhysRevLett.128.237401} {\bibfield
  {journal} {\bibinfo  {journal} {Phys. Rev. Lett.}\ }\textbf {\bibinfo
  {volume} {128}},\ \bibinfo {pages} {237401} (\bibinfo {year}
  {2022})}\BibitemShut {NoStop}%
\bibitem [{\citenamefont {Chernikov}\ \emph {et~al.}(2014)\citenamefont
  {Chernikov}, \citenamefont {Berkelbach}, \citenamefont {Hill}, \citenamefont
  {Rigosi}, \citenamefont {Li}, \citenamefont {Aslan}, \citenamefont
  {Reichman}, \citenamefont {Hybertsen},\ and\ \citenamefont
  {Heinz}}]{chernikov_Exciton_2014}%
  \BibitemOpen
  \bibfield  {author} {\bibinfo {author} {\bibfnamefont {A.}~\bibnamefont
  {Chernikov}}, \bibinfo {author} {\bibfnamefont {T.~C.}\ \bibnamefont
  {Berkelbach}}, \bibinfo {author} {\bibfnamefont {H.~M.}\ \bibnamefont
  {Hill}}, \bibinfo {author} {\bibfnamefont {A.}~\bibnamefont {Rigosi}},
  \bibinfo {author} {\bibfnamefont {Y.}~\bibnamefont {Li}}, \bibinfo {author}
  {\bibfnamefont {O.~B.}\ \bibnamefont {Aslan}}, \bibinfo {author}
  {\bibfnamefont {D.~R.}\ \bibnamefont {Reichman}}, \bibinfo {author}
  {\bibfnamefont {M.~S.}\ \bibnamefont {Hybertsen}},\ and\ \bibinfo {author}
  {\bibfnamefont {T.~F.}\ \bibnamefont {Heinz}},\ }\href
  {https://doi.org/10.1103/PhysRevLett.113.076802} {\bibfield  {journal}
  {\bibinfo  {journal} {Phys. Rev. Lett.}\ }\textbf {\bibinfo {volume} {113}},\
  \bibinfo {pages} {076802} (\bibinfo {year} {2014})}\BibitemShut {NoStop}%
\bibitem [{\citenamefont {Sun}\ \emph {et~al.}(2022)\citenamefont {Sun},
  \citenamefont {Zhu}, \citenamefont {Qin}, \citenamefont {Liu}, \citenamefont
  {Tang}, \citenamefont {L{\"u}}, \citenamefont {Rahman}, \citenamefont
  {Yildirim},\ and\ \citenamefont {Lu}}]{sun_Enhanced_2022}%
  \BibitemOpen
  \bibfield  {author} {\bibinfo {author} {\bibfnamefont {X.}~\bibnamefont
  {Sun}}, \bibinfo {author} {\bibfnamefont {Y.}~\bibnamefont {Zhu}}, \bibinfo
  {author} {\bibfnamefont {H.}~\bibnamefont {Qin}}, \bibinfo {author}
  {\bibfnamefont {B.}~\bibnamefont {Liu}}, \bibinfo {author} {\bibfnamefont
  {Y.}~\bibnamefont {Tang}}, \bibinfo {author} {\bibfnamefont {T.}~\bibnamefont
  {L{\"u}}}, \bibinfo {author} {\bibfnamefont {S.}~\bibnamefont {Rahman}},
  \bibinfo {author} {\bibfnamefont {T.}~\bibnamefont {Yildirim}},\ and\
  \bibinfo {author} {\bibfnamefont {Y.}~\bibnamefont {Lu}},\ }\href
  {https://doi.org/10.1038/s41586-022-05193-z} {\bibfield  {journal} {\bibinfo
  {journal} {Nature}\ }\textbf {\bibinfo {volume} {610}},\ \bibinfo {pages}
  {478} (\bibinfo {year} {2022})}\BibitemShut {NoStop}%
\bibitem{Supplement_Material} See Supplemental Material at [url], which includes 
Refs. [28, 29, 33, 35, 40-45, 49, 50], for the brief derivation of Eq. 1,
expressions of Bogoliubov excitation spectrum, calculation of BEC critical temperature with interaction, discussion on the effect of screening and calculation of vortex energy.
\bibitem [{\citenamefont {Ridolfi}\ \emph {et~al.}(2018)\citenamefont
	{Ridolfi}, \citenamefont {Lewenkopf},\ and\ \citenamefont
	{Pereira}}]{ridolfi_Excitonic_2018}%
\BibitemOpen
\bibfield  {author} {\bibinfo {author} {\bibfnamefont {E.}~\bibnamefont
		{Ridolfi}}, \bibinfo {author} {\bibfnamefont {C.~H.}\ \bibnamefont
		{Lewenkopf}},\ and\ \bibinfo {author} {\bibfnamefont {V.~M.}\ \bibnamefont
		{Pereira}},\ }\href {https://doi.org/10.1103/PhysRevB.97.205409} {\bibfield
	{journal} {\bibinfo  {journal} {Phys. Rev. B}\ }\textbf {\bibinfo {volume}
		{97}},\ \bibinfo {pages} {205409} (\bibinfo {year} {2018})}\BibitemShut
{NoStop}%
\bibitem [{\citenamefont {Simbulan}\ \emph {et~al.}(2021)\citenamefont
	{Simbulan}, \citenamefont {Huang}, \citenamefont {Peng}, \citenamefont {Li},
	\citenamefont {Gomez~Sanchez}, \citenamefont {Lin}, \citenamefont {Lu},
	\citenamefont {Yang}, \citenamefont {Qi}, \citenamefont {Cheng},
	\citenamefont {Lu},\ and\ \citenamefont {Lan}}]{simbulan_Selective_2021}%
\BibitemOpen
\bibfield  {author} {\bibinfo {author} {\bibfnamefont {K.~B.}\ \bibnamefont
		{Simbulan}}, \bibinfo {author} {\bibfnamefont {T.-D.}\ \bibnamefont {Huang}},
	\bibinfo {author} {\bibfnamefont {G.-H.}\ \bibnamefont {Peng}}, \bibinfo
	{author} {\bibfnamefont {F.}~\bibnamefont {Li}}, \bibinfo {author}
	{\bibfnamefont {O.~J.}\ \bibnamefont {Gomez~Sanchez}}, \bibinfo {author}
	{\bibfnamefont {J.-D.}\ \bibnamefont {Lin}}, \bibinfo {author} {\bibfnamefont
		{C.-I.}\ \bibnamefont {Lu}}, \bibinfo {author} {\bibfnamefont {C.-S.}\
		\bibnamefont {Yang}}, \bibinfo {author} {\bibfnamefont {J.}~\bibnamefont
		{Qi}}, \bibinfo {author} {\bibfnamefont {S.-J.}\ \bibnamefont {Cheng}},
	\bibinfo {author} {\bibfnamefont {T.-H.}\ \bibnamefont {Lu}},\ and\ \bibinfo
	{author} {\bibfnamefont {Y.-W.}\ \bibnamefont {Lan}},\ }\href
{https://doi.org/10.1021/acsnano.0c10823} {\bibfield  {journal} {\bibinfo
		{journal} {ACS Nano}\ }\textbf {\bibinfo {volume} {15}},\ \bibinfo {pages}
	{3481} (\bibinfo {year} {2021})}\BibitemShut {NoStop}%
\bibitem [{\citenamefont {Aivazian}\ \emph {et~al.}(2015)\citenamefont
  {Aivazian}, \citenamefont {Gong}, \citenamefont {Jones}, \citenamefont {Chu},
  \citenamefont {Yan}, \citenamefont {Mandrus}, \citenamefont {Zhang},
  \citenamefont {Cobden}, \citenamefont {Yao},\ and\ \citenamefont
  {Xu}}]{aivazian_Magnetic_2015}%
  \BibitemOpen
  \bibfield  {author} {\bibinfo {author} {\bibfnamefont {G.}~\bibnamefont
  {Aivazian}}, \bibinfo {author} {\bibfnamefont {Z.}~\bibnamefont {Gong}},
  \bibinfo {author} {\bibfnamefont {A.~M.}\ \bibnamefont {Jones}}, \bibinfo
  {author} {\bibfnamefont {R.-L.}\ \bibnamefont {Chu}}, \bibinfo {author}
  {\bibfnamefont {J.}~\bibnamefont {Yan}}, \bibinfo {author} {\bibfnamefont
  {D.~G.}\ \bibnamefont {Mandrus}}, \bibinfo {author} {\bibfnamefont
  {C.}~\bibnamefont {Zhang}}, \bibinfo {author} {\bibfnamefont
  {D.}~\bibnamefont {Cobden}}, \bibinfo {author} {\bibfnamefont
  {W.}~\bibnamefont {Yao}},\ and\ \bibinfo {author} {\bibfnamefont
  {X.}~\bibnamefont {Xu}},\ }\href {https://doi.org/10.1038/nphys3201}
  {\bibfield  {journal} {\bibinfo  {journal} {Nat. Phys.}\ }\textbf {\bibinfo
  {volume} {11}},\ \bibinfo {pages} {148} (\bibinfo {year} {2015})}\BibitemShut
  {NoStop}%
\bibitem [{\citenamefont {Stier}\ \emph {et~al.}(2016)\citenamefont {Stier},
  \citenamefont {McCreary}, \citenamefont {Jonker}, \citenamefont {Kono},\ and\
  \citenamefont {Crooker}}]{stier_Exciton_2016}%
  \BibitemOpen
  \bibfield  {author} {\bibinfo {author} {\bibfnamefont {A.~V.}\ \bibnamefont
  {Stier}}, \bibinfo {author} {\bibfnamefont {K.~M.}\ \bibnamefont {McCreary}},
  \bibinfo {author} {\bibfnamefont {B.~T.}\ \bibnamefont {Jonker}}, \bibinfo
  {author} {\bibfnamefont {J.}~\bibnamefont {Kono}},\ and\ \bibinfo {author}
  {\bibfnamefont {S.~A.}\ \bibnamefont {Crooker}},\ }\href
  {https://doi.org/10.1038/ncomms10643} {\bibfield  {journal} {\bibinfo
  {journal} {Nat. Commun.}\ }\textbf {\bibinfo {volume} {7}},\ \bibinfo {pages}
  {10643} (\bibinfo {year} {2016})}\BibitemShut {NoStop}%
\bibitem [{\citenamefont {Kim}\ \emph {et~al.}(2014)\citenamefont {Kim},
  \citenamefont {Hong}, \citenamefont {Jin}, \citenamefont {Shi}, \citenamefont
  {Chang}, \citenamefont {Chiu}, \citenamefont {Li},\ and\ \citenamefont
  {Wang}}]{kim_Ultrafast_2014}%
  \BibitemOpen
  \bibfield  {author} {\bibinfo {author} {\bibfnamefont {J.}~\bibnamefont
  {Kim}}, \bibinfo {author} {\bibfnamefont {X.}~\bibnamefont {Hong}}, \bibinfo
  {author} {\bibfnamefont {C.}~\bibnamefont {Jin}}, \bibinfo {author}
  {\bibfnamefont {S.-F.}\ \bibnamefont {Shi}}, \bibinfo {author} {\bibfnamefont
  {C.-Y.~S.}\ \bibnamefont {Chang}}, \bibinfo {author} {\bibfnamefont {M.-H.}\
  \bibnamefont {Chiu}}, \bibinfo {author} {\bibfnamefont {L.-J.}\ \bibnamefont
  {Li}},\ and\ \bibinfo {author} {\bibfnamefont {F.}~\bibnamefont {Wang}},\
  }\href {https://doi.org/10.1126/science.1258122} {\bibfield  {journal}
  {\bibinfo  {journal} {Science}\ }\textbf {\bibinfo {volume} {346}},\ \bibinfo
  {pages} {1205} (\bibinfo {year} {2014})}\BibitemShut {NoStop}%
\bibitem [{\citenamefont {Sie}\ \emph {et~al.}(2015)\citenamefont {Sie},
  \citenamefont {McIver}, \citenamefont {Lee}, \citenamefont {Fu},
  \citenamefont {Kong},\ and\ \citenamefont
  {Gedik}}]{sie_Valleyselective_2015}%
  \BibitemOpen
  \bibfield  {author} {\bibinfo {author} {\bibfnamefont {E.~J.}\ \bibnamefont
  {Sie}}, \bibinfo {author} {\bibfnamefont {J.~W.}\ \bibnamefont {McIver}},
  \bibinfo {author} {\bibfnamefont {Y.-H.}\ \bibnamefont {Lee}}, \bibinfo
  {author} {\bibfnamefont {L.}~\bibnamefont {Fu}}, \bibinfo {author}
  {\bibfnamefont {J.}~\bibnamefont {Kong}},\ and\ \bibinfo {author}
  {\bibfnamefont {N.}~\bibnamefont {Gedik}},\ }\href
  {https://doi.org/10.1038/nmat4156} {\bibfield  {journal} {\bibinfo  {journal}
  {Nat. Mater.}\ }\textbf {\bibinfo {volume} {14}},\ \bibinfo {pages} {290}
  (\bibinfo {year} {2015})}\BibitemShut {NoStop}%
\bibitem [{\citenamefont {Shahnazaryan}\ \emph {et~al.}(2017)\citenamefont
  {Shahnazaryan}, \citenamefont {Iorsh}, \citenamefont {Shelykh},\ and\
  \citenamefont {Kyriienko}}]{shahnazaryan_Excitonexciton_2017}%
  \BibitemOpen
  \bibfield  {author} {\bibinfo {author} {\bibfnamefont {V.}~\bibnamefont
  {Shahnazaryan}}, \bibinfo {author} {\bibfnamefont {I.}~\bibnamefont {Iorsh}},
  \bibinfo {author} {\bibfnamefont {I.~A.}\ \bibnamefont {Shelykh}},\ and\
  \bibinfo {author} {\bibfnamefont {O.}~\bibnamefont {Kyriienko}},\ }\href
  {https://doi.org/10.1103/PhysRevB.96.115409} {\bibfield  {journal} {\bibinfo
  {journal} {Phys. Rev. B}\ }\textbf {\bibinfo {volume} {96}},\ \bibinfo
  {pages} {115409} (\bibinfo {year} {2017})}\BibitemShut {NoStop}%
\bibitem [{\citenamefont {Erkensten}\ \emph {et~al.}(2021)\citenamefont
  {Erkensten}, \citenamefont {Brem},\ and\ \citenamefont
  {Malic}}]{erkensten_Excitonexciton_2021}%
  \BibitemOpen
  \bibfield  {author} {\bibinfo {author} {\bibfnamefont {D.}~\bibnamefont
  {Erkensten}}, \bibinfo {author} {\bibfnamefont {S.}~\bibnamefont {Brem}},\
  and\ \bibinfo {author} {\bibfnamefont {E.}~\bibnamefont {Malic}},\ }\href
  {https://doi.org/10.1103/PhysRevB.103.045426} {\bibfield  {journal} {\bibinfo
   {journal} {Phys. Rev. B}\ }\textbf {\bibinfo {volume} {103}},\ \bibinfo
  {pages} {045426} (\bibinfo {year} {2021})}\BibitemShut {NoStop}%
\bibitem [{\citenamefont {Stier}\ \emph {et~al.}(2018)\citenamefont {Stier},
  \citenamefont {Wilson}, \citenamefont {Velizhanin}, \citenamefont {Kono},
  \citenamefont {Xu},\ and\ \citenamefont
  {Crooker}}]{stier_Magnetooptics_2018}%
  \BibitemOpen
  \bibfield  {author} {\bibinfo {author} {\bibfnamefont {A.~V.}\ \bibnamefont
  {Stier}}, \bibinfo {author} {\bibfnamefont {N.~P.}\ \bibnamefont {Wilson}},
  \bibinfo {author} {\bibfnamefont {K.~A.}\ \bibnamefont {Velizhanin}},
  \bibinfo {author} {\bibfnamefont {J.}~\bibnamefont {Kono}}, \bibinfo {author}
  {\bibfnamefont {X.}~\bibnamefont {Xu}},\ and\ \bibinfo {author}
  {\bibfnamefont {S.~A.}\ \bibnamefont {Crooker}},\ }\href
  {https://doi.org/10.1103/PhysRevLett.120.057405} {\bibfield  {journal}
  {\bibinfo  {journal} {Phys. Rev. Lett.}\ }\textbf {\bibinfo {volume} {120}},\
  \bibinfo {pages} {057405} (\bibinfo {year} {2018})}\BibitemShut {NoStop}%
\bibitem [{\citenamefont {Liu}\ \emph {et~al.}(2019)\citenamefont {Liu},
  \citenamefont {{van Baren}}, \citenamefont {Taniguchi}, \citenamefont
  {Watanabe}, \citenamefont {Chang},\ and\ \citenamefont
  {Lui}}]{liu_Magnetophotoluminescence_2019}%
  \BibitemOpen
  \bibfield  {author} {\bibinfo {author} {\bibfnamefont {E.}~\bibnamefont
  {Liu}}, \bibinfo {author} {\bibfnamefont {J.}~\bibnamefont {{van Baren}}},
  \bibinfo {author} {\bibfnamefont {T.}~\bibnamefont {Taniguchi}}, \bibinfo
  {author} {\bibfnamefont {K.}~\bibnamefont {Watanabe}}, \bibinfo {author}
  {\bibfnamefont {Y.-C.}\ \bibnamefont {Chang}},\ and\ \bibinfo {author}
  {\bibfnamefont {C.~H.}\ \bibnamefont {Lui}},\ }\href
  {https://doi.org/10.1103/PhysRevB.99.205420} {\bibfield  {journal} {\bibinfo
  {journal} {Phys. Rev. B}\ }\textbf {\bibinfo {volume} {99}},\ \bibinfo
  {pages} {205420} (\bibinfo {year} {2019})}\BibitemShut {NoStop}%
\bibitem [{\citenamefont {Goryca}\ \emph {et~al.}(2019)\citenamefont {Goryca},
  \citenamefont {Li}, \citenamefont {Stier}, \citenamefont {Taniguchi},
  \citenamefont {Watanabe}, \citenamefont {Courtade}, \citenamefont {Shree},
  \citenamefont {Robert}, \citenamefont {Urbaszek}, \citenamefont {Marie},\
  and\ \citenamefont {Crooker}}]{goryca_Revealing_2019}%
  \BibitemOpen
  \bibfield  {author} {\bibinfo {author} {\bibfnamefont {M.}~\bibnamefont
  {Goryca}}, \bibinfo {author} {\bibfnamefont {J.}~\bibnamefont {Li}}, \bibinfo
  {author} {\bibfnamefont {A.~V.}\ \bibnamefont {Stier}}, \bibinfo {author}
  {\bibfnamefont {T.}~\bibnamefont {Taniguchi}}, \bibinfo {author}
  {\bibfnamefont {K.}~\bibnamefont {Watanabe}}, \bibinfo {author}
  {\bibfnamefont {E.}~\bibnamefont {Courtade}}, \bibinfo {author}
  {\bibfnamefont {S.}~\bibnamefont {Shree}}, \bibinfo {author} {\bibfnamefont
  {C.}~\bibnamefont {Robert}}, \bibinfo {author} {\bibfnamefont
  {B.}~\bibnamefont {Urbaszek}}, \bibinfo {author} {\bibfnamefont
  {X.}~\bibnamefont {Marie}},\ and\ \bibinfo {author} {\bibfnamefont {S.~A.}\
  \bibnamefont {Crooker}},\ }\href {https://doi.org/10.1038/s41467-019-12180-y}
  {\bibfield  {journal} {\bibinfo  {journal} {Nat. Commun.}\ }\textbf {\bibinfo
  {volume} {10}},\ \bibinfo {pages} {4172} (\bibinfo {year}
  {2019})}\BibitemShut {NoStop}%
\bibitem [{\citenamefont {Pethick}\ and\ \citenamefont
  {Smith}(2008)}]{pethick_Bose_2008}%
  \BibitemOpen
  \bibfield  {author} {\bibinfo {author} {\bibfnamefont {C.~J.}\ \bibnamefont
  {Pethick}}\ and\ \bibinfo {author} {\bibfnamefont {H.}~\bibnamefont
  {Smith}},\ }\href {https://doi.org/10.1017/CBO9780511802850} {\emph {\bibinfo
  {title} {Bose\textendash{{Einstein Condensation}} in {{Dilute Gases}}}}},\
  \bibinfo {edition} {2nd}\ ed.\ (\bibinfo {address} {{Cambridge}},\ \bibinfo
  {year} {2008})\BibitemShut {NoStop}%
\bibitem [{\citenamefont {{Ben-Tabou de-Leon}}\ and\ \citenamefont
  {Laikhtman}(2001)}]{ben-taboude-leon_Excitonexciton_2001}%
  \BibitemOpen
  \bibfield  {author} {\bibinfo {author} {\bibfnamefont {S.}~\bibnamefont
  {{Ben-Tabou de-Leon}}}\ and\ \bibinfo {author} {\bibfnamefont
  {B.}~\bibnamefont {Laikhtman}},\ }\href
  {https://doi.org/10.1103/PhysRevB.63.125306} {\bibfield  {journal} {\bibinfo
  {journal} {Phys. Rev. B}\ }\textbf {\bibinfo {volume} {63}},\ \bibinfo
  {pages} {125306} (\bibinfo {year} {2001})}\BibitemShut {NoStop}%
\bibitem [{\citenamefont {Griffin}(1996)}]{griffin_Conserving_1996}%
  \BibitemOpen
  \bibfield  {author} {\bibinfo {author} {\bibfnamefont {A.}~\bibnamefont
  {Griffin}},\ }\href {https://doi.org/10.1103/PhysRevB.53.9341} {\bibfield
  {journal} {\bibinfo  {journal} {Phys. Rev. B}\ }\textbf {\bibinfo {volume}
  {53}},\ \bibinfo {pages} {9341} (\bibinfo {year} {1996})}\BibitemShut
  {NoStop}%
\bibitem [{\citenamefont {Hutchinson}\ \emph {et~al.}(1997)\citenamefont
  {Hutchinson}, \citenamefont {Zaremba},\ and\ \citenamefont
  {Griffin}}]{hutchinson_Finite_1997}%
  \BibitemOpen
  \bibfield  {author} {\bibinfo {author} {\bibfnamefont {D.~A.~W.}\
  \bibnamefont {Hutchinson}}, \bibinfo {author} {\bibfnamefont
  {E.}~\bibnamefont {Zaremba}},\ and\ \bibinfo {author} {\bibfnamefont
  {A.}~\bibnamefont {Griffin}},\ }\href
  {https://doi.org/10.1103/PhysRevLett.78.1842} {\bibfield  {journal} {\bibinfo
   {journal} {Phys. Rev. Lett.}\ }\textbf {\bibinfo {volume} {78}},\ \bibinfo
  {pages} {1842} (\bibinfo {year} {1997})}\BibitemShut {NoStop}%
\bibitem [{\citenamefont {Shi}\ and\ \citenamefont
  {Griffin}(1998)}]{shi_Finitetemperature_1998}%
  \BibitemOpen
  \bibfield  {author} {\bibinfo {author} {\bibfnamefont {H.}~\bibnamefont
  {Shi}}\ and\ \bibinfo {author} {\bibfnamefont {A.}~\bibnamefont {Griffin}},\
  }\href {https://doi.org/10.1016/S0370-1573(98)00015-5} {\bibfield  {journal}
  {\bibinfo  {journal} {Phys. Rep.}\ }\textbf {\bibinfo {volume} {304}},\
  \bibinfo {pages} {1} (\bibinfo {year} {1998})}\BibitemShut {NoStop}%
\bibitem [{\citenamefont {Cataudella}\ and\ \citenamefont
  {Minnhagen}(1990)}]{cataudella_Simple_1990}%
  \BibitemOpen
  \bibfield  {author} {\bibinfo {author} {\bibfnamefont {V.}~\bibnamefont
  {Cataudella}}\ and\ \bibinfo {author} {\bibfnamefont {P.}~\bibnamefont
  {Minnhagen}},\ }\href {https://doi.org/10.1016/0921-4534(90)90042-D}
  {\bibfield  {journal} {\bibinfo  {journal} {Physica C}\ }\textbf {\bibinfo
  {volume} {166}},\ \bibinfo {pages} {442} (\bibinfo {year}
  {1990})}\BibitemShut {NoStop}%
\bibitem [{\citenamefont {Jensen}\ and\ \citenamefont
  {Minnhagen}(1991)}]{jensen_Twodimensional_1991}%
  \BibitemOpen
  \bibfield  {author} {\bibinfo {author} {\bibfnamefont {H.~J.}\ \bibnamefont
  {Jensen}}\ and\ \bibinfo {author} {\bibfnamefont {P.}~\bibnamefont
  {Minnhagen}},\ }\href {https://doi.org/10.1103/PhysRevLett.66.1630}
  {\bibfield  {journal} {\bibinfo  {journal} {Phys. Rev. Lett.}\ }\textbf
  {\bibinfo {volume} {66}},\ \bibinfo {pages} {1630} (\bibinfo {year}
  {1991})}\BibitemShut {NoStop}%
\bibitem [{\citenamefont {Miu}\ \emph {et~al.}(1998)\citenamefont {Miu},
  \citenamefont {Jakob}, \citenamefont {Haibach}, \citenamefont {Kluge},
  \citenamefont {Frey}, \citenamefont {{Voss-de Haan}},\ and\ \citenamefont
  {Adrian}}]{miu_Lengthscaledependent_1998}%
  \BibitemOpen
  \bibfield  {author} {\bibinfo {author} {\bibfnamefont {L.}~\bibnamefont
  {Miu}}, \bibinfo {author} {\bibfnamefont {G.}~\bibnamefont {Jakob}}, \bibinfo
  {author} {\bibfnamefont {P.}~\bibnamefont {Haibach}}, \bibinfo {author}
  {\bibfnamefont {{\relax Th}.}~\bibnamefont {Kluge}}, \bibinfo {author}
  {\bibfnamefont {U.}~\bibnamefont {Frey}}, \bibinfo {author} {\bibfnamefont
  {P.}~\bibnamefont {{Voss-de Haan}}},\ and\ \bibinfo {author} {\bibfnamefont
  {H.}~\bibnamefont {Adrian}},\ }\href
  {https://doi.org/10.1103/PhysRevB.57.3144} {\bibfield  {journal} {\bibinfo
  {journal} {Phys. Rev. B}\ }\textbf {\bibinfo {volume} {57}},\ \bibinfo
  {pages} {3144} (\bibinfo {year} {1998})}\BibitemShut {NoStop}%
\bibitem [{\citenamefont {Babaev}(2004)}]{babaev_Phase_2004}%
  \BibitemOpen
  \bibfield  {author} {\bibinfo {author} {\bibfnamefont {E.}~\bibnamefont
  {Babaev}},\ }\href {https://doi.org/10.1016/j.nuclphysb.2004.02.021}
  {\bibfield  {journal} {\bibinfo  {journal} {Nucl. Phys. B}\ }\textbf
  {\bibinfo {volume} {686}},\ \bibinfo {pages} {397} (\bibinfo {year}
  {2004})}\BibitemShut {NoStop}%
\bibitem [{\citenamefont {Kobayashi}\ \emph {et~al.}(2019)\citenamefont
  {Kobayashi}, \citenamefont {Eto},\ and\ \citenamefont
  {Nitta}}]{kobayashi_BerezinskiiKosterlitzThouless_2019}%
  \BibitemOpen
  \bibfield  {author} {\bibinfo {author} {\bibfnamefont {M.}~\bibnamefont
  {Kobayashi}}, \bibinfo {author} {\bibfnamefont {M.}~\bibnamefont {Eto}},\
  and\ \bibinfo {author} {\bibfnamefont {M.}~\bibnamefont {Nitta}},\ }\href
  {https://doi.org/10.1103/PhysRevLett.123.075303} {\bibfield  {journal}
  {\bibinfo  {journal} {Phys. Rev. Lett.}\ }\textbf {\bibinfo {volume} {123}},\
  \bibinfo {pages} {075303} (\bibinfo {year} {2019})}\BibitemShut {NoStop}%
\bibitem [{\citenamefont {Fetter}(2009)}]{fetter_Rotating_2009}%
  \BibitemOpen
  \bibfield  {author} {\bibinfo {author} {\bibfnamefont {A.~L.}\ \bibnamefont
  {Fetter}},\ }\href {https://doi.org/10.1103/RevModPhys.81.647} {\bibfield
  {journal} {\bibinfo  {journal} {Rev. Mod. Phys.}\ }\textbf {\bibinfo {volume}
  {81}},\ \bibinfo {pages} {647} (\bibinfo {year} {2009})}\BibitemShut
  {NoStop}%
\bibitem [{\citenamefont {Nelson}\ and\ \citenamefont
  {Kosterlitz}(1977)}]{nelson_Universal_1977}%
  \BibitemOpen
  \bibfield  {author} {\bibinfo {author} {\bibfnamefont {D.~R.}\ \bibnamefont
  {Nelson}}\ and\ \bibinfo {author} {\bibfnamefont {J.~M.}\ \bibnamefont
  {Kosterlitz}},\ }\href {https://doi.org/10.1103/PhysRevLett.39.1201}
  {\bibfield  {journal} {\bibinfo  {journal} {Phys. Rev. Lett.}\ }\textbf
  {\bibinfo {volume} {39}},\ \bibinfo {pages} {1201} (\bibinfo {year}
  {1977})}\BibitemShut {NoStop}%
\bibitem [{\citenamefont {Prokof'ev}\ and\ \citenamefont
  {Svistunov}(2001)}]{prokofev_Worm_2001}%
  \BibitemOpen
  \bibfield  {author} {\bibinfo {author} {\bibfnamefont {N.}~\bibnamefont
  {Prokof'ev}}\ and\ \bibinfo {author} {\bibfnamefont {B.}~\bibnamefont
  {Svistunov}},\ }\href {https://doi.org/10.1103/PhysRevLett.87.160601}
  {\bibfield  {journal} {\bibinfo  {journal} {Phys. Rev. Lett.}\ }\textbf
  {\bibinfo {volume} {87}},\ \bibinfo {pages} {160601} (\bibinfo {year}
  {2001})}\BibitemShut {NoStop}%
\bibitem [{\citenamefont {Prokof'ev}\ and\ \citenamefont
  {Svistunov}(2002)}]{prokofev_Twodimensional_2002}%
  \BibitemOpen
  \bibfield  {author} {\bibinfo {author} {\bibfnamefont {N.}~\bibnamefont
  {Prokof'ev}}\ and\ \bibinfo {author} {\bibfnamefont {B.}~\bibnamefont
  {Svistunov}},\ }\href {https://doi.org/10.1103/PhysRevA.66.043608} {\bibfield
   {journal} {\bibinfo  {journal} {Phys. Rev. A}\ }\textbf {\bibinfo {volume}
  {66}},\ \bibinfo {pages} {043608} (\bibinfo {year} {2002})}\BibitemShut
  {NoStop}%
\end{thebibliography}
\end{document}